\documentclass[12pt]{iopart}
\usepackage{graphicx}
\begin{document}

\title[Laser interactions with nonlinear transparent media]{Laser dynamics in nonlinear transparent media with electron plasma generation: effects of electron-hole radiative recombinations}

\author{P. Kameni Nteutse$^1$, Alain M. Dikand\'e$^2$ and 
S. Zekeng$^1$}

\address{$^1$ Laboratory of Mechanics, Materials and Structures, Department of Physics, Faculty of Science, University of Yaounde I P.O. Box 812 Yaounde, Cameroon.}
\address{$^2$Laboratory of Research on Advanced Materials and Nonlinear Science (LaRAMaNS), Department of Physics, Faculty of Science, University of Buea P.O. Box 63 Buea, Cameroon.}
\ead{dikande.alain@ubuea.cm}
\vspace{10pt}
\begin{indented}
\item[Resubmission date: ]January 2021
\end{indented}

\begin{abstract}
The performance of optical devices manufactured via laser micromachining on nonlinear transparent materials, relies usually on three main factors which are the characteristic laser parameters (i.e. the laser power, pulse duration and pulse repetition rate), characteristic properties of host materials (e.g. their chromatic dispersions, optical nonlinearities or self-focusing features, etc.) and the relative importance of physical processes such as the avalanche impact ionization, multiphoton ionization and electron-hole radiative recombination processes. These factors act in conjunction to impose the regime of laser operation, in particular their competition determines the appropriate laser operation regime. In this work a theoretical study is proposed to explore the effects of the competition between multiphoton absorption, plasma ionization and electron-hole radiative recombination processes, on the laser dynamics in transparent materials with Kerr nonlinearity. The study rests on a model consisting of a K-order nonlinear complex Ginzburg-Landau equation, coupled to a first-order equation describing time variation of the electron plasma density. An analysis of stability of continuous waves, following the modulational-instability approach, reveals that the combination of multiphoton absorption and electron-hole radiative recombination processes can be detrimental or favorable to continuous-wave operation, depending on the group-velocity dispersion of the host medium. Numerical simulations of the model equations in the full nonlinear regime, reveal the eixstence of pulse trains the amplitudes of which are enhanced by the radiative recombination processes. Numerical results for the density of the induced electron plasma feature two distinct regimes of time evolution, depending on the strength of the electron-hole radiative recombination processes.  
\end{abstract}

%

\section{Introduction}
Femtosecond laser micromachining nowadays offers the most reliable and portable tool in a broad range of modern industrial material processing \cite{r1,r2,r3,r4}, its applications extend from accurate manufacturing of electronic devices, to fine drilling and machining of hard metals, ceramics and soft plastics into various micro textures for improvement of functions and properties of end products \cite{r2,r3,r4,r5}. In these applications laser pulses focused on a dielectric medium are absorbed via nonlinear photo-ionization mechanism \cite{r6,r7}, leading to a permanent modification of material structure at scales on the order of nanometers. In the specific context of transparent materials \cite{r6,r7,r8}, at low pulse powers the modification will be a smooth refractive index change which can be exploited advantageously in the fabrication of photonic devices \cite{r2,r9}. However, at higher pulse powers the modification gives rise to more complex processes such as birefringences, periodic nanoplanes aligning themselves orthogonally to the laser polarization to form periodic nanogratings, change in the electronic structure due to electron and hole productions from charge ionization with the generation of electron plasma, electron-hole radiative recombination processes \cite{r2,r6,r9}, etc..  
\par In accordance with the fineness required for the end product, optical fields used in laser material processing can be grouped in two categories \cite{r2,r9,r10}: continuous-wave (CW) lasers, which usually extend up to several kilowatts, and pulsed lasers with an average power spanning well below one kilowatt thus providing a wide range of wavelengths and pulse duration, as well as pulse repetition rates \cite{r10}. Due to these attributes, pulse lasers can allow micromachining with high resolution in depth and therefore offer a rich potential for applications in drilling \cite{r2},
cutting \cite{r11}, welding \cite{r12,r13}, ablation \cite{r12},
material surface texturing and scripting \cite{r12,r13}. We remark that besides
their short duration and high powers, femtosecond pulse lasers have been most attractive owing to their minimal thermal drawbacks \cite{r14,r15}. Indeed femtosecond lasers are able to accumulate heat such as to minimize defect-induced damages, as a matter of fact this heat accumulation prevents undesired physical casualties as for instance the formation of microcracks, material bending, etc. \cite{r16} notably during laser processing involving transparent materials \cite{r17,r18,r19}. 
\par Theoretical investigations of femtosecond laser processing on transparent materials have attracted a little attention in the past \cite{r7,r19a,r19b}. Yet theoretical studies are useful for they provide fundamental knowledge relative to the global picture of the laser dynamics, relevant for a good understanding of its distinct possible operation regimes in the prospect of an optimization of the micromachining technology. Indeed, since femtosecond lasers are optical fields with duration far below picosecond, they belong to a specific class of lasers known as ultrashort lasers \cite{r20}. Lasers in this specific class operate typically in pulsed modes of relatively high powers, nevertheless in some contexts they can be tailored to operate in the CW regime \cite{r2,r9}. This is for instance the case when their input powers are below the typical power of a high-intensity optical pulse, or when the input field is of low power and is designed to grow upon propagation from CW mode to a high-intensity
pulse. Such growth can be regarded as an instability-induced dynamical transition of the CW laser, this instability is best known as modulational-instability process \cite{r21,r22,r23,r24} and involves
CW breakup into high-power optical fields, leading ultimately to a
pulse via a regime dominated by pulse-train structures \cite{r25,r26,r27,r28}.
\par It follows that for a good understanding of laser processing we must have a clear picture of possible dynamical regimes of the laser. In nonlinear optical materials this issue can be understood in terms of laser self-starting dynamics, where the input is assumed to be a CW field whose amplitude grows upon propagation until a threshold
amplitude. Beyond this threshold amplitude the CW mode becomes modulational unstable,
typically this instability will first generate weakly nonlinear pulse trains
which decay subsequently into high-intensity temporal pulses.
\par In this work we examine the dynamics of a model of femtosecond laser interacting with Kerr nonlinear media, taking into consideration material modification and the generation of an electron plasma. In refs. \cite{r7,r19a,r19b}, mathematical models were proposed to describe femtosecond laser interaction with Kerr nonlinear optical media. These models are generally composed of a complex Ginzburg–Landau (CGL) equation for laser
propagation, coupled to a first-order time evolution equation for the electron plasma density. Multiphoton ionization result in a K-order nonlinear term both in the CGL equation and in the rate equation. In these previous works \cite{r7,r19a,r19b} the authors discussed only thermal aspects of laser interactions with Kerr-type transparent materials, in this respect they established that multiphoton ionization processes would drastically minimize thermal drawbacks. 
\par The model considered in the present study emanes from two previous models, namely the model proposed in ref. \cite{r19b} where account was taken of the contribution of avalanche ionization due to K-photon absorption processes to the electron plasma generation, and the model of ref. \cite{r19a} where electron-recombination processes were taken into consideration. In our case we extend the model to the physical context in which electron-hole radiative recombination processes are relevant. This translates into an additional term in the time evolution equation for the plasma density, quadratic in the density variable and competing with the avalanche impact ionization and the multi-photon ionization processes. \\
To start we analyze the system dynamics in the CW regime. In this respect we find steady-state solutions, and explore their stability by means of the modulational-instability theory. A global stability picture will be proposed in terms of a two-dimensional complex parameter space, mapped by the real and imaginary parts of the coefficient of spatial amplification of noise over a finite range of values of the modulation frequency. Next we discuss the system dynamics in the full nonlinear regime, starting with fixed-point solutions and then carrying out numerical simulations, using a sixth-order Runge-Kutta algorithm adopted from Luther \cite{r30}, to generate nonlinear time series of the laser amplitude and the electron plasma density, for different values of the radiative recombination coefficient and of the photon number $K$. 
\section{The model, CW solutions and modulational instability}
 Consider an optical field propagating along the $z$ axis of a transparent medium with Kerr nonlinearity. We assume that the energy stored by the propagating laser modifies the electronic structure of the bulk material, creating an electron plasma of variable density and multiphoton ionization processes. Instructively the problem was addressed experimentally \cite{r7,r10,r19a,r19b} in the contexts of femtosecond laser pulses focused on silica materials. It was established that for a strong focusing geometry the femtosecond laser can cause bulk damages in the optical material, followed by a narrow track with submicron width indicating a filamentary propagation of the pulse laser. \par In refs. \cite{r7,r10,r19a,r19b} numerical simulations were carried out to investigate the distribution of the energy stored in the bulk material by the propagating laser. While the laser equation was exactly similar in all these previous studies \cite{r7,r10,r19a,r19b}, the time-evolution equations for the electron plasma density were different. The difference resided in the nature of physical processes assumed to contribute to the electron plasma ionization in each of these studies, as a matter of fact the work of ref. \cite{r7} was mainly interested in the impact of a linear recombination process on the time evolution of the electron plasma density, whereas in refs. \cite{r10,r19a,r19b} the linerar recombination process was neglected in favor of radiative recombinations only \cite{r10}, avalanche impact only \cite{r19a} or a combination of avalanche impact ionization and linear recombination processes \cite{r19b}. However in none of these studies a detailed analysis of the system dynamics was considered. \par
 In the present study we pay interest to the system dynamics, considering a model that combines the effects of avalanche ionization process, the electron-hole radiative recombination and multiphoton ionization processes. The model can be represented by a propagation equation for the laser field given in terms of the cubic complex Ginzburg-Landau equation with a $K$-order nonlinear term \cite{r7,r10,r19a,r19b}, coupled to a first-order ordinary differential equation accounting for time evolution of the electron plasma density, i.e.:
\begin{eqnarray}
	i\frac{\partial u}{\partial z}&=&\delta\frac{\partial^2u}{\partial t^2}-\sigma|u|^2u-i\gamma(1-i\omega_0\tau_0)\rho u-i\mu|u|^{2K-2}u, \label{j1}\\
	\frac{\partial \rho}{\partial t}&=&\nu |u|^2 \rho + \alpha |u|^{2K}-a\rho^{2}. \label{j2}
\end{eqnarray}

Instructively, we assumed a paraxial approximation for the field propagation considering the propagation axis to be $z$, and ignored beam diffractions in the plane transverse to the propagation axis. The first term in the right-hand side of eq. (\ref{j1}) accounts for the group-velocity dispersion, the second term accounts for the intrinsic (i.e. Kerr) nonlinearity of the host material, the third term describes both the plasma absorption and laser defocusing, and the last term describes multiphoton absorption processes with $\mu$ the $K$-photon ionization rate. \par 
Eq. (\ref{j2}) describes the temporal evolution of the electron plasma density $\rho$, we remark that this equation differs from those considered in refs. \cite{r7,r19a,r19b} where the same problem was discussed. Indeed eq. (\ref{j2}) takes into account all together the contributions from avalanche impact ionization (first term), multiphoton ionization (second term) and electron-hole radiative recombination (last term). This remark also holds for the study carried out in ref. \cite{r29}, where electron-hole radiative recombinations were not taken into consideration. Parameters in eq. (\ref{j1}) and eq. (\ref{j2}) have the following physical meanings and characteristic units (where necessary) \cite{r7,r10,r19a,r19b}:
\begin{itemize}
 \item $\delta$ {\bf (in fs$^2/m$)} is the coefficient of group-velocity dispersion,
 \item $\sigma$ {\bf (in $m^2/W$)} is the Kerr nonlinearity coefficient,
 \item $\omega_0$ {\bf (in MHz) and $\tau_0$ (in fs)} are the frequency and characteristic relaxation time of the electron plasma respectively,
 \item $\gamma$ {\bf (in $m^2$)} is the cross section for inverse Bremsstrahlung,
 \item $\nu$ {\bf (in $m^2/eV$)} is the coefficient of avalanche impact ionization,
 \item $\alpha$ is the multiphoton ionization coefficient in the dense medium,
 \item and $a$ {\bf (in $m^3/s$)} is the electron-hole radiative recombination coefficient.
\end{itemize}
Typical values for these parameters are found in most experimental works dealing with the problem, as for instance in the study of femtosecond laser filamentations in transpartent media \cite{r10}, laser micro-modification of fused silica \cite{r7} and so on. However, for a theoretical study such as the present one, experimental values of these parameters are of no useful given that the mathematical model, represented by eqs. (\ref{j1}) and (\ref{j2}), involves normalized variables such as the propagation time $t$, the propagation distance $z$, the field amplitude $u$ as well as some coefficients in these two equations \cite{r10,r7}. Therefore in our study we shall select arbitrary but reasonable values for characteristic parameters of the model, but keeping track of their appropriate signs. \par 
Being two nonlinear equations, general solutions to the coupled set eqs. (\ref{j1})-(\ref{j2}) are nonlinear waves. Nevertheless, provided specific conditions, linear solutions including harmonic waves and CWs can also exist for the same set. Thus steady-state CW solutions to eqs. (\ref{j1})-(\ref{j2}) can be expressed:
\begin{equation}
	u(z)=\sqrt{I_{p}}\exp\left(iP_{c} z\right), \quad \rho=\rho_{0}, \label{i1} 
\end{equation}
which upon substitution in (\ref{j1})-(\ref{j2}) give:
\begin{eqnarray}
	P_{c}&=&\sigma I_{p}+\gamma\omega_0\tau_0\rho_{0}, \qquad I_{p}= \left(\frac{\sigma}{\mu\omega_0\tau_0}\right)^{\frac{1}{K-2}}, \label{j6} \\
  \rho_{0}&=&\frac{\nu}{2a}I_{p}\left[1-\sqrt{1+\frac{4a\alpha\sigma}{\nu^2\mu\omega_0\tau_0}} \right]. \label{j7} 
\end{eqnarray}
Here $P_{c}$, the CW wave number, is fixed by the input power $I_{p}=\vert u\vert^2$ as well as the equilibrium value $\rho_{0}$ of the electron plasma density $\rho$. Given that these two last quantities (i.e. $I_{p}$ and $\rho_{0}$) depend on characteristic parameters of the model, they cannot be arbitrary and hence can be tuned by varying characteristic parameters of the model. \par
To look into the stability of the CW solutions eqs. (\ref{i1}), we carry out a modulational-instability analysis assuming a small-amplitude perturbation $f(z,t)$ of the CW amplitude $\sqrt{I_{p}}$. Steady-state solutions to eqs. (\ref{j1})-(\ref{j2}) thus become: 
\begin{equation}
	u(z,t)=\left[u_{0}+f(z,t)\right]\exp(iP_{c}z), \qquad
	\rho(t)=\rho_{0}+\delta\rho(t). \label{j9} 
\end{equation}
Replacing in eqs. (\ref{j1})-(\ref{j2}) and linearizing, we obtain:
\begin{eqnarray}
	i\frac{\partial f}{\partial z}&-&\delta\frac{\partial^2 f}{\partial t^2}+C^{(1)}_{K}\,(f+f^{*})=-i\gamma(1-i\omega_0\tau_0)u_{0}\delta\rho(t), \label{j10}\\\nonumber\\
	\frac{\partial \delta\rho(t)}{\partial t}&-&q_0\delta\rho(t)=C^{(2)}_{K}\,(f+f^{*}), \label{j11} 
\end{eqnarray}

Where,
\begin{eqnarray}
	C^{(1)}_K&=&\sigma u^2_0 + i\mu (K-1) u^{2K-2}_0 ,\quad C^{(2)}_K=\frac{1}{\gamma}(\alpha\gamma K-\mu \nu)u^{2K-1}_0,  \nonumber \\
q_0&=&\nu u^2_{0}+\frac{2a\mu}{\gamma}u^{2K-2}_{0}, \label{j12}
\end{eqnarray} 
$f^{*}$ in the linear equations (\ref{j10}) and (\ref{j11}) denotes the complex conjugate of $f$. The first-order inhomogeneous linear equation (\ref{j11}) can be solved by means of Green's function technique \cite{r23,r29}, yielding: 
\begin{eqnarray}
	\delta\rho(t)=C^{(2)}_{K}\int_{-\infty}^{t}(f+f^{*})e^{-q_0(t'-t)}dt'. \label{j13}
\end{eqnarray} 

Because of the presence of $f^{*}$ in eq. (\ref{j10}), we must consider its complex conjugate. With this consideration, as solution to eq. (\ref{j10}) we pick $f(z,t)=A_{1}\exp\left(\kappa z+i\Omega t\right)$ and $f^{*}(z,t)=A_{2}\exp\left(\kappa z+i\Omega t\right)$, where $\kappa$ is the coefficient of spatial amplification of the perturbation and $\Omega$ is the modulation frequency for the perturbation. These solutions lead to the following matrix-form eigenvalue problem:

\begin{eqnarray}
	\kappa
	 \left(\begin{array}{c}
		A_{1}\\ A_{2}
	\end{array} \right) &=&\left[
\left(\begin{array}{cc}		
         M_{1} & M_{2}\\
		M^{*}_{2} & M^{*}_{1}
	\end{array}\right) - S_{0} \left(
                        \begin{array}{cc}
		1&1\\
		1 & 1
	\end{array}\right)\right]\left(
                        \begin{array}{cc}
                       A_1 \\ A_2
                        \end{array}
                        \right) \nonumber \\	
                        &-&T_{0}\left( \begin{array}{cc}
		N_{1}&N_{1}\\
		N^{*}_{1} & N^{*}_{1}
	\end{array}\right) \left(\begin{array}{c}
		A_{1}\\
		A_{2}
	\end{array}\right), \label{21}
\end{eqnarray}	
where:
\begin{eqnarray}
 M_{1}&=&i(\delta\Omega^2+\sigma u_{0}^2), \quad M_{2}=i\sigma u_{0}^2, \label{22} \\
	S_{0}&=&\mu(K-1)u_{0}^{2K-2}, \quad N_{1}=1-i\omega_0\tau_0, \label{23}\\
	T_{0}&=&(K\alpha\gamma-\mu\nu)u^{2K}_{0}\left(i\Omega-\nu u^2_{0}-2a\mu u^{2K-2}_0/\gamma\right)^{-1}. \label{24}
\end{eqnarray} 
The two possible eigenvalues of the above $2\times 2$ matrix equation are:
\begin{eqnarray}
	\kappa_{1,2}&=& -\mu(K-1)u^{2K-2}_0\left[1+\frac{\gamma(K\alpha\gamma-\mu\nu)u_{0}^{2}}{\mu(K-1)(i\Omega\gamma-\nu u_{0}^2\gamma-2a\mu  u_{0}^{2K-2})}\right] \nonumber \\
	&\pm& \sqrt{(S_{0}+T_{0})^2-(\delta\Omega^2+\sigma u_{0}^2)^2+\sigma^2 u_{0}^4-2T_{0}\delta\Omega^2\omega_0\tau_{0}}, \label{25} 
\end{eqnarray} 
where subscripts $1,2$ refer to the plus (+) and minus (-) signs respectively. From the standpoint of modulational-instability theory, CW stability will depend on the sign and nature (i.e. real or complex) of the two eigenvalues $\kappa_{1,2}$. Most generally the following possible situations are expected:
\begin{itemize}
\item When the real part of $\kappa$ is zero, the CW solution will be always stable irrespective of the sign of its imaginary part.
\item 
When the real part of $\kappa$ is negative, the CW solution will be asymptotically stable (i.e. is stabilized after some roundtrips) irrespective of the sign of its imaginary part.
\item When the real part of $\kappa$ is positive, the CW regime will be always unstable.
\end{itemize}

\begin{figure}\centering
\begin{minipage}[c]{0.5\textwidth}
\includegraphics[width=2.8in, height= 2.5in]{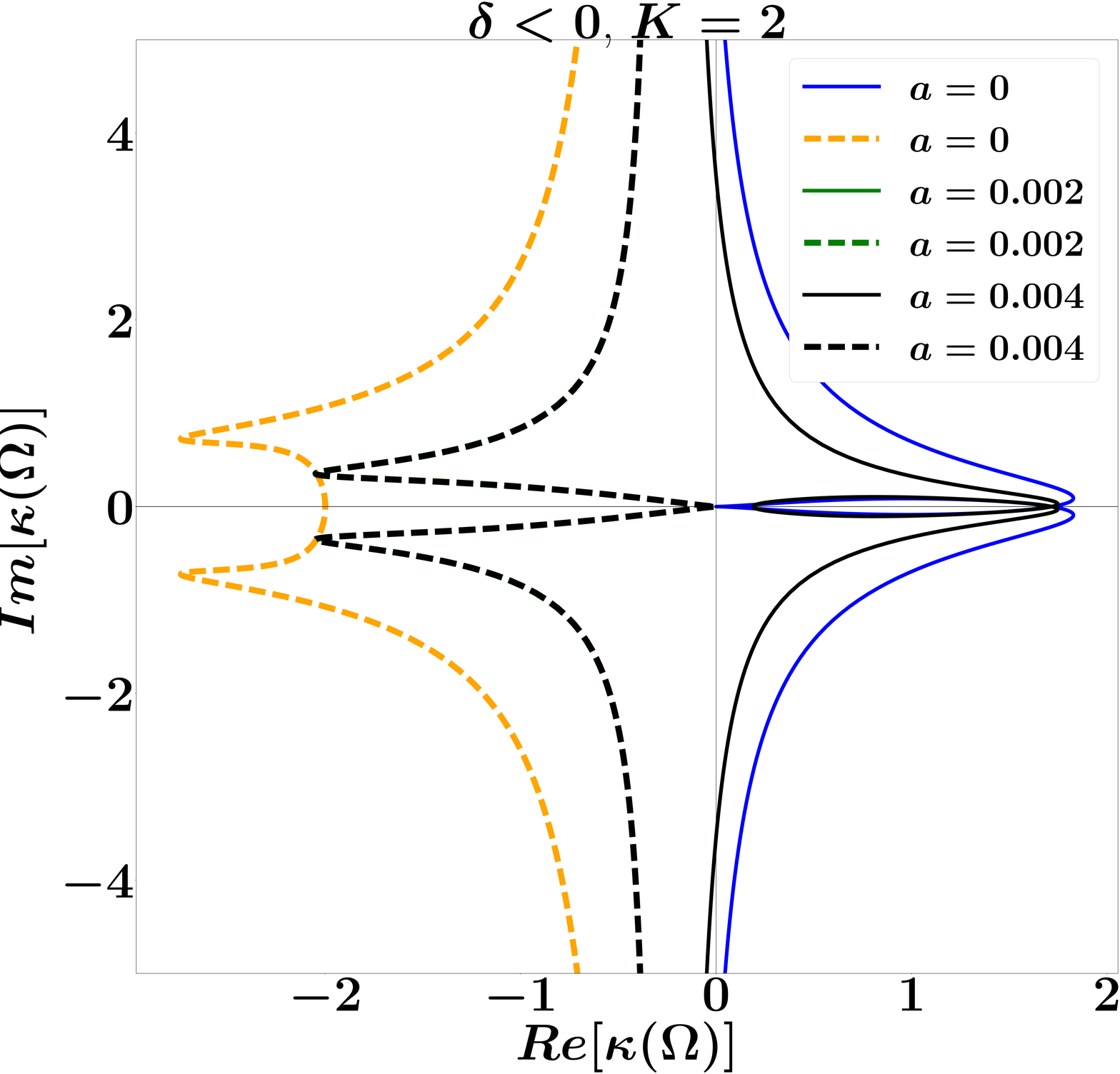}
\end{minipage}%
\begin{minipage}[c]{0.5\textwidth}
\includegraphics[width=2.8in, height= 2.5in]{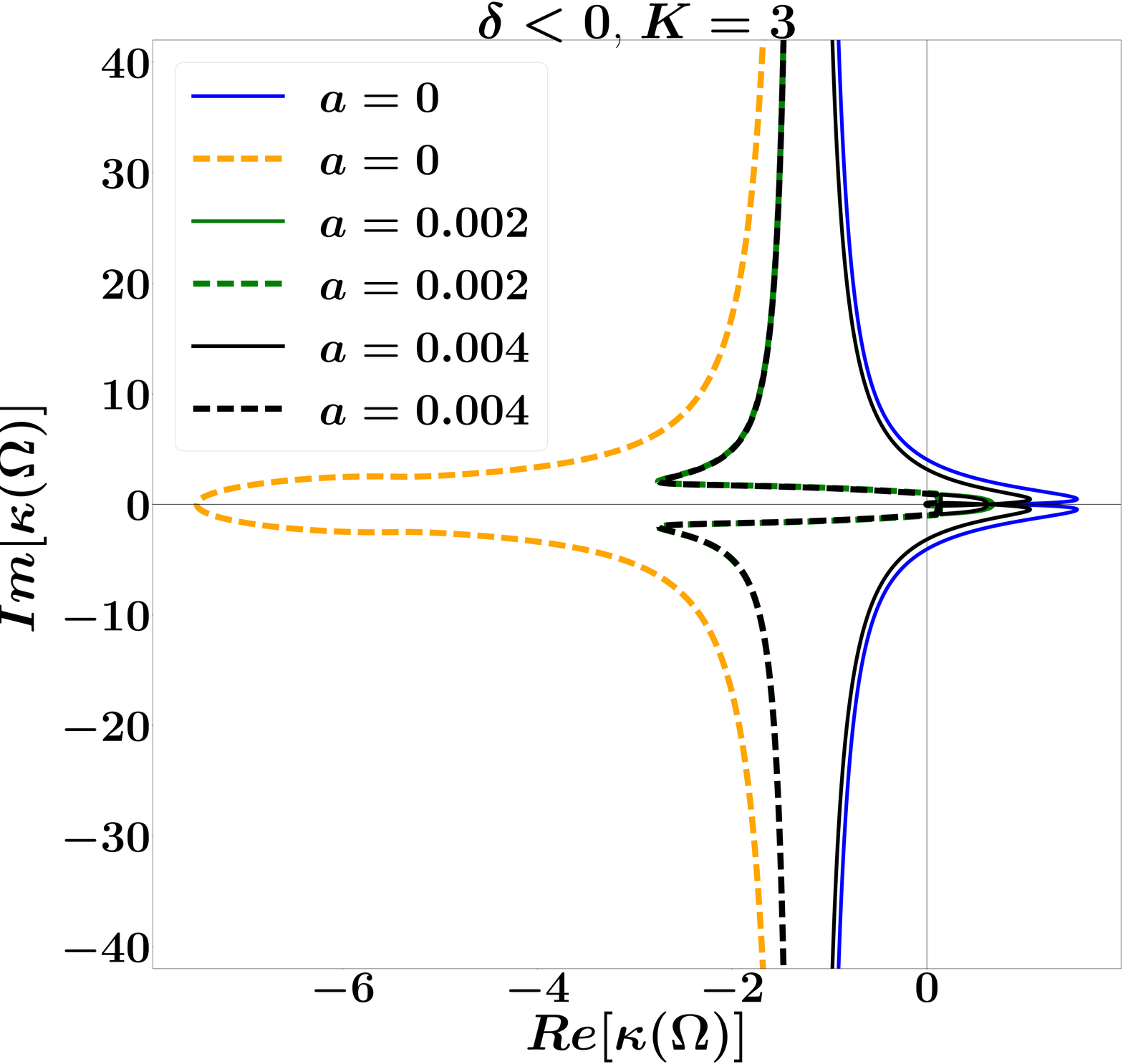}
\end{minipage}\\
\vspace{0.5truecm}
\begin{minipage}[c]{0.5\textwidth}
\includegraphics[width=2.8in, height= 2.5in]{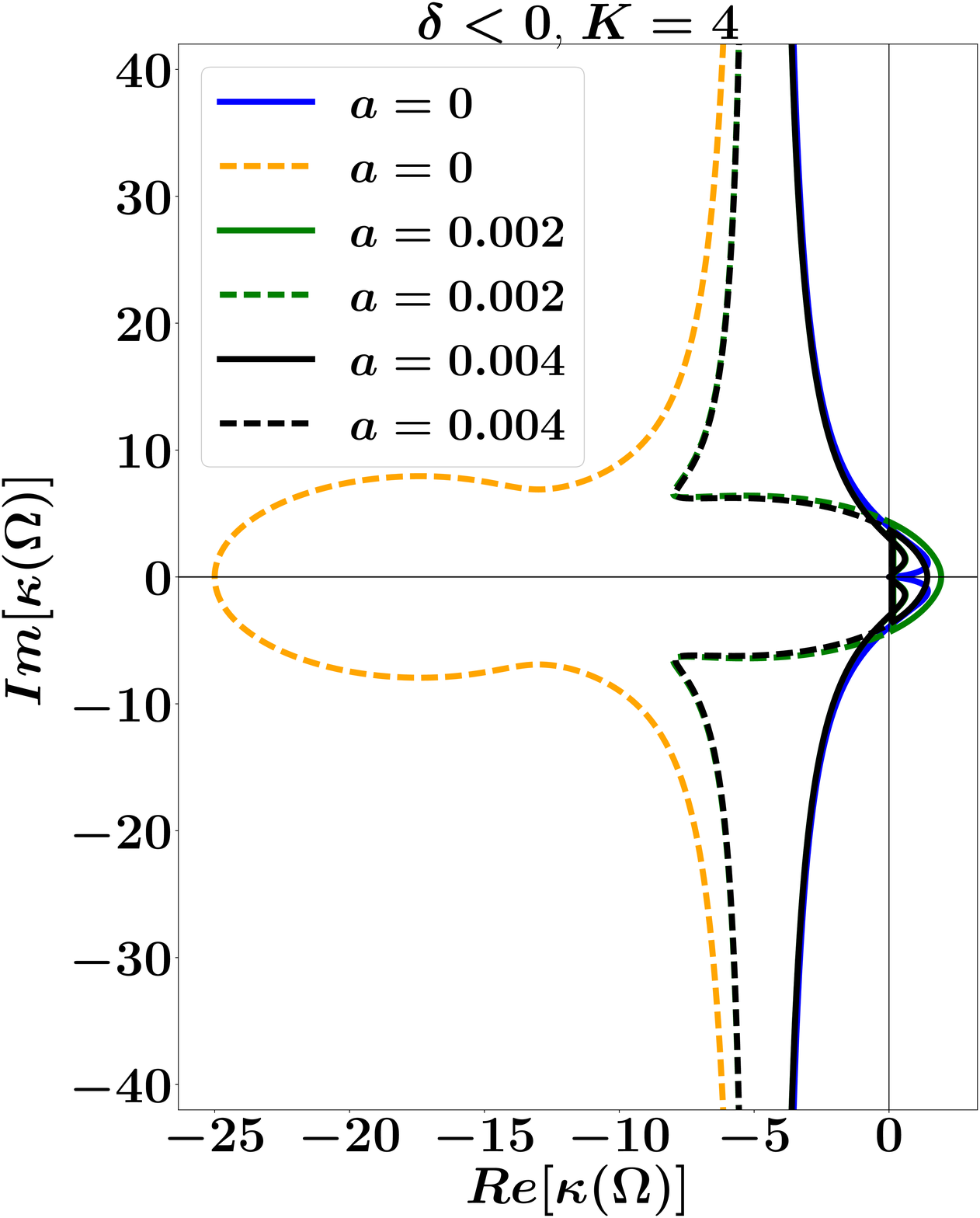}
\end{minipage}%
\begin{minipage}[c]{0.5\textwidth}
\includegraphics[width=2.8in, height= 2.5in]{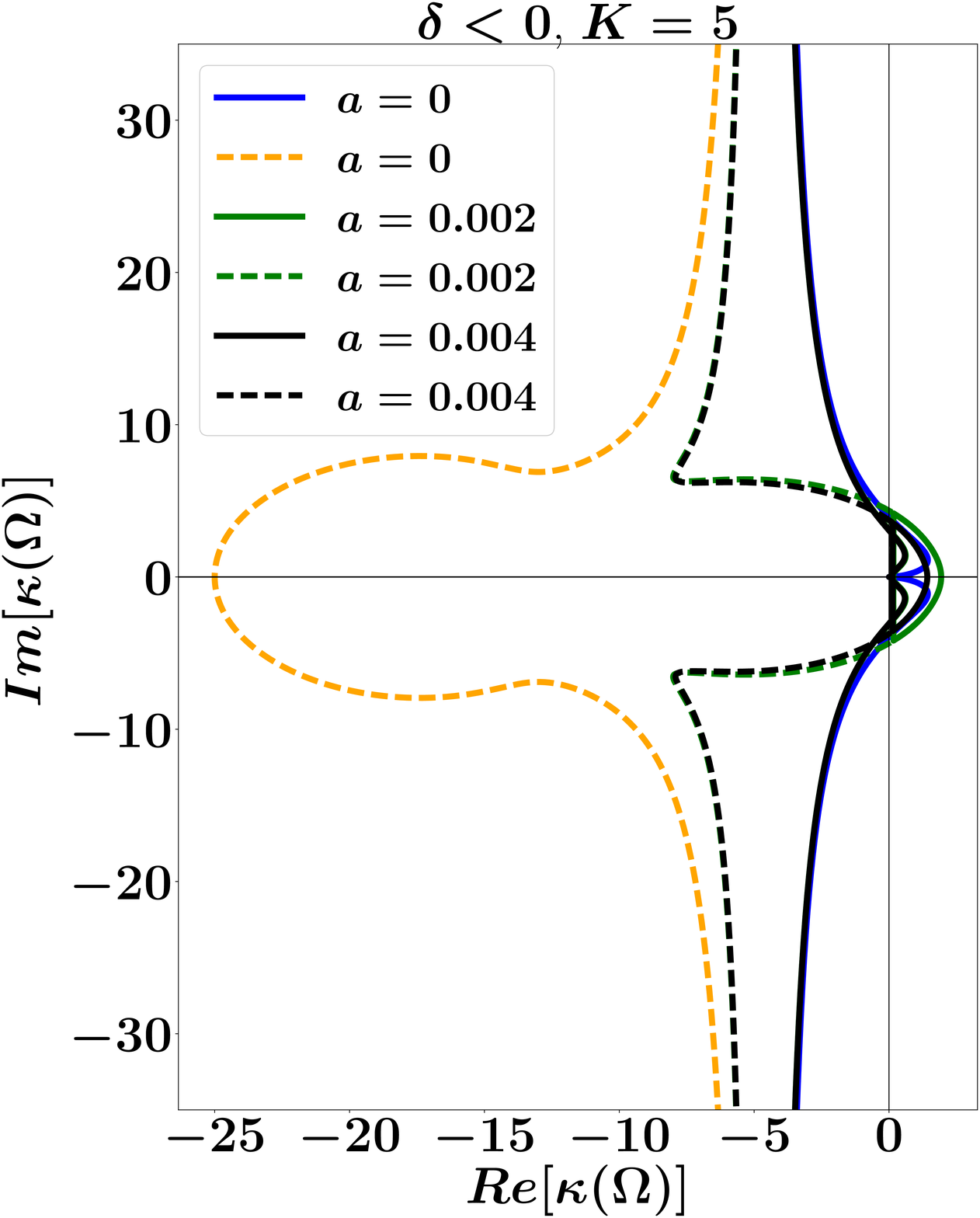}
\end{minipage}
\caption{(Color online) Imaginary versus real parts of $\kappa_1$ (full curve) and $\kappa_2$ (dashed curve) for $K=2, 3, 4, 5$. The radiative recombination coefficient $a$ is varied as $a$= $0$, $0.001$, $0.002$, $0.003$, $0.004$. $\alpha=0.6$ ,$\nu=0.5$, $\mu=0.1$, $I_p=2.5$, $\omega_0\tau_0=0.2$, $\sigma=0.8$, $\delta=-0.5$, $\gamma=0.1$.}{\label{fig1}}
  \end{figure}

\begin{figure}\centering
\begin{minipage}[c]{0.5\textwidth}
\includegraphics[width=2.8in, height= 2.5in]{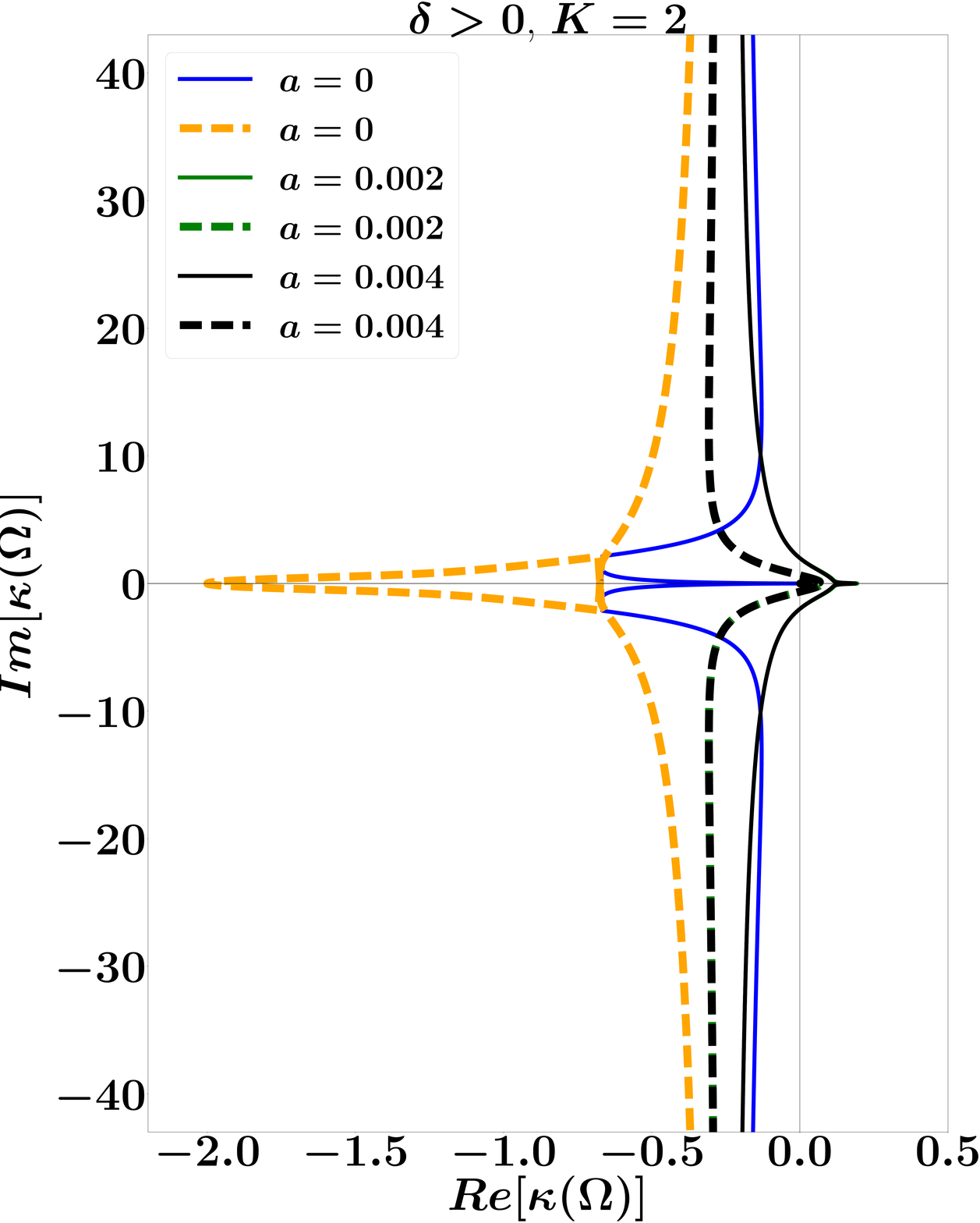}
\end{minipage}%
\begin{minipage}[c]{0.5\textwidth}
\includegraphics[width=2.8in, height= 2.5in]{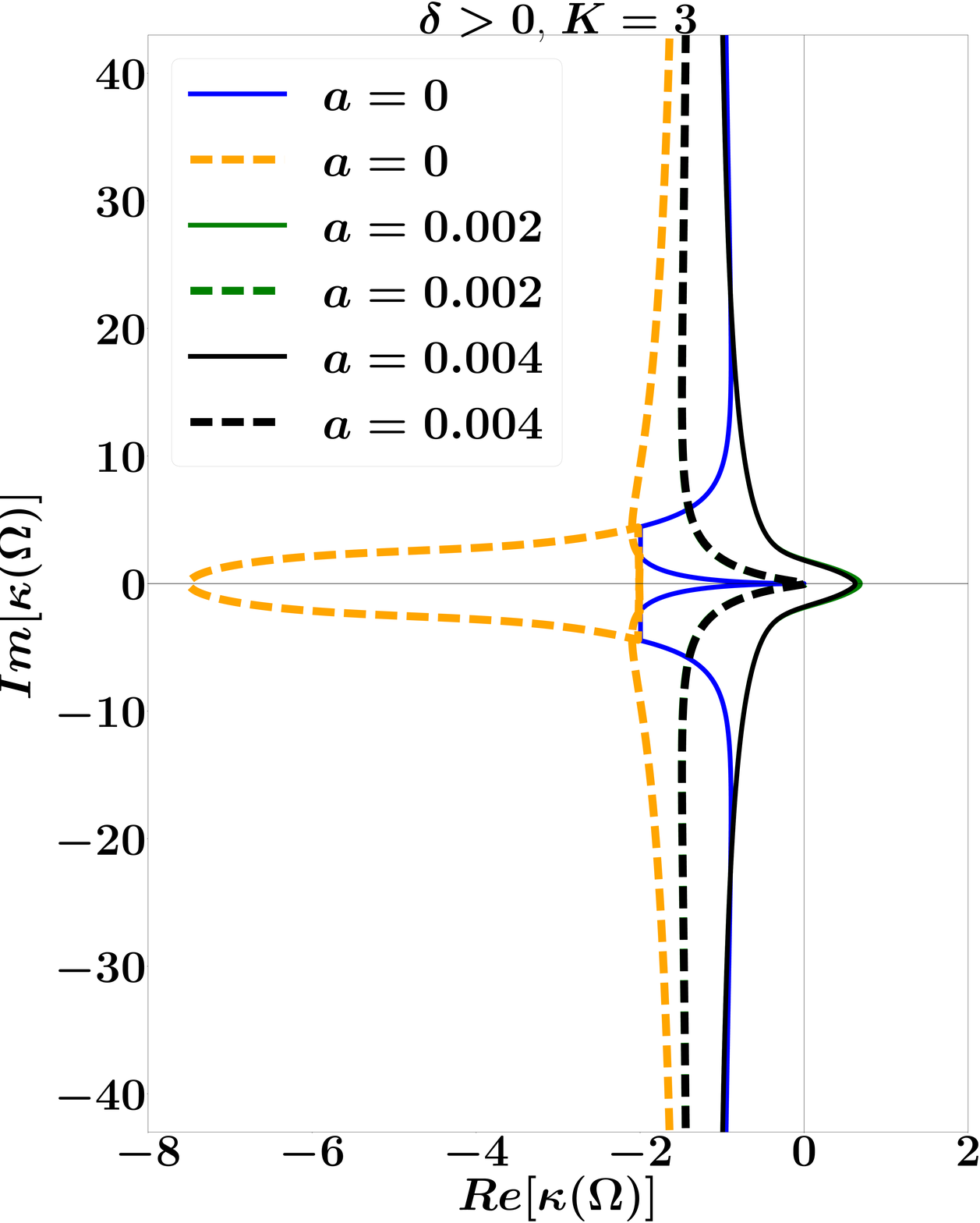}
\end{minipage}\\
\vspace{0.5truecm}
\begin{minipage}[c]{0.5\textwidth}
\includegraphics[width=2.8in, height= 2.5in]{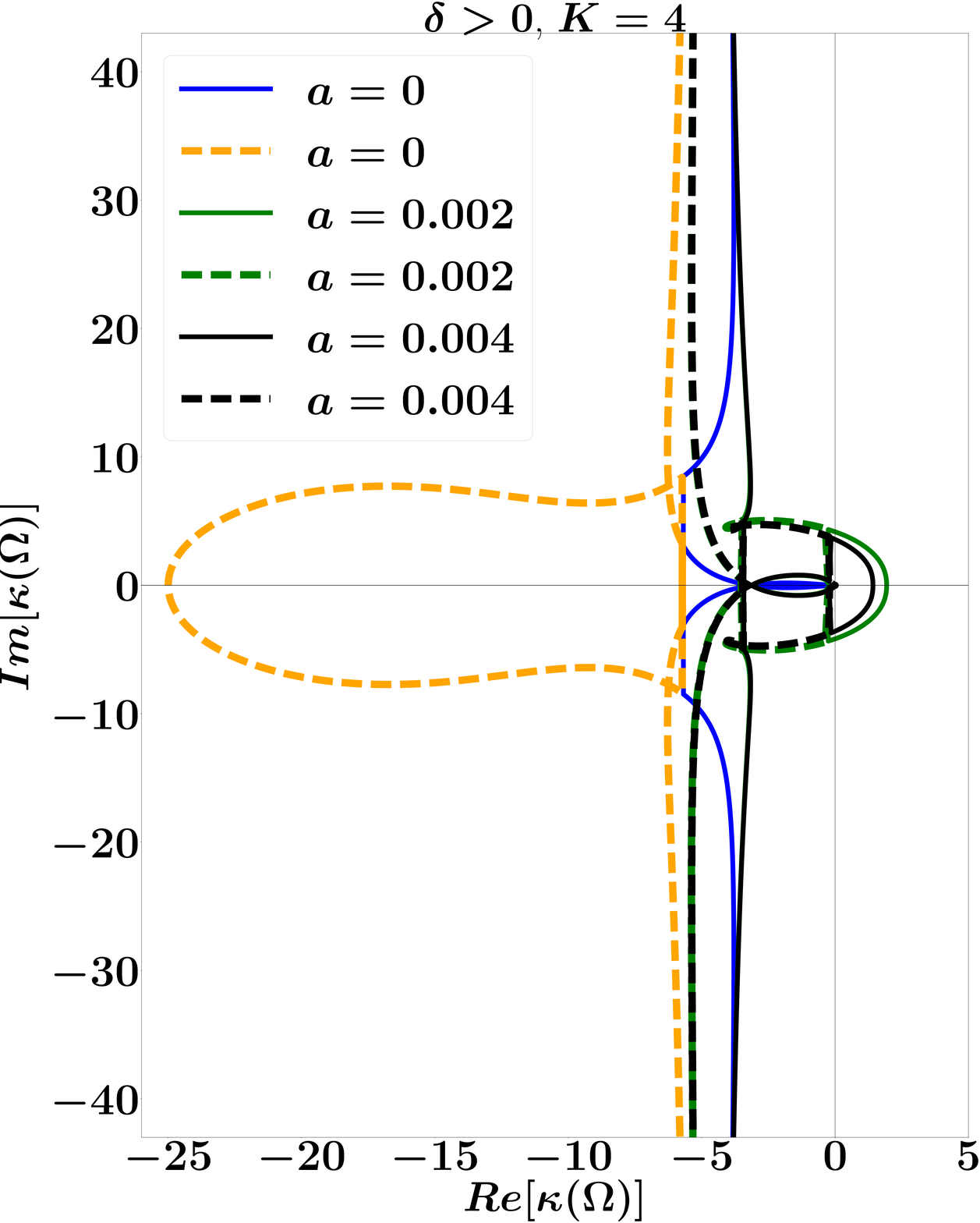}
\end{minipage}%
\begin{minipage}[c]{0.5\textwidth}
\includegraphics[width=2.8in, height= 2.5in]{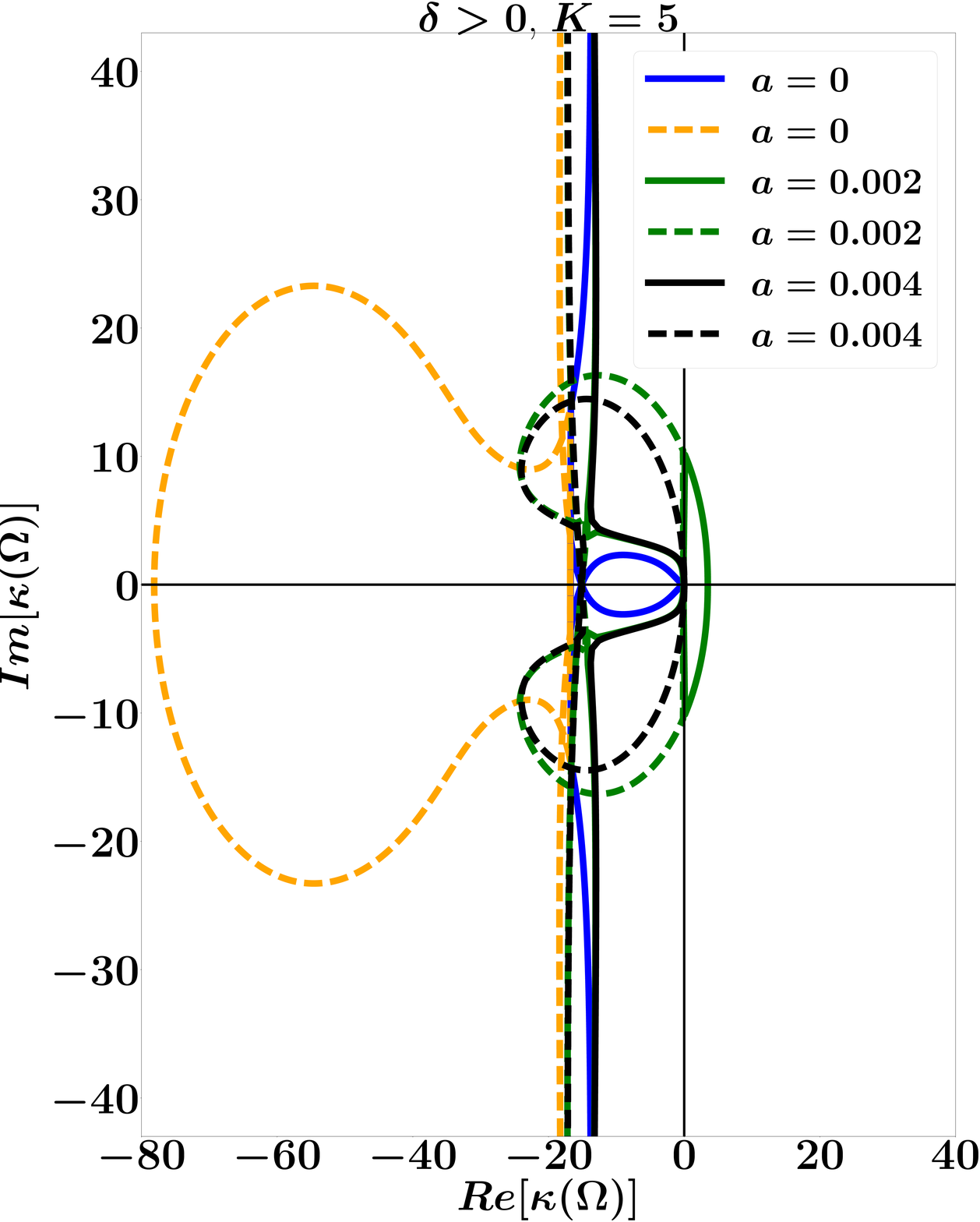}
\end{minipage}
      \caption{(Color online) Imaginary versus real parts of $\kappa_1$ (full curve) and $\kappa_2$ (dashed curve) for $K=2,3,4,5$. The radiative recombination coefficient $a$ is varied as $a$= $0$, $0.001$, $0.002$, $0.003$, $0.004$. $\alpha=0.6$ ,$\nu=0.5$, $\mu=0.1$, $I_p=2.5$, $\omega_0\tau_0=0.2$, $\sigma=0.8$, $\delta=0.5$, $\gamma=0.1$.}{\label{fig2}}
  \end{figure}

Given that the two eigenvalues are functions of the modulation frequency $\Omega$, we find it more appropriate to first consider the CW stability at zero modulation frequency. In this later case the eigenvalues are:

\begin{equation}
\kappa_{1}=0, \qquad \kappa_{2}=-2\mu(K-1)u^{2K-2}_{0}+\frac{2\gamma(K\alpha\gamma-\mu\nu)u_{0}^{2K}}{\nu u_{0}^2\gamma+2 a\mu u^{2K-2}_{0}}. \label{18}
\end{equation}
It turns out that laser self-starting (i.e. CW instability) will be favored provided $\kappa_{2}>0$, or in terms of formula (\ref{18});
\begin{equation}
a<\frac{K\gamma(\alpha\gamma-\mu\nu)}{2\mu^2(K-1)I^{K-2}_{p}}.\label{ss}
\end{equation}
Quantitatively, this condition implies two possible characteristic values of the radiative recombination coefficient $a$ above which laser self-starting can occur: One is negative for $\alpha\gamma<\mu\nu$ and hence is nonphysical, whereas the positive and physical one is conditioned by $\alpha\gamma>\mu\nu$ and is:
\begin{eqnarray}
	a_{th}=\frac{K\gamma(\alpha\gamma-\mu\nu)}{2\mu^2(K-1)I^{K-2}_{p}}. \label{19}
\end{eqnarray}
In concrete terms the quantity $a_{th}$ sets a threshold value of the electron-ion recombination coefficient, above which the laser will self-start.\\
Due to the strong dependence of $\kappa_{1,2}$ in eq. (\ref{25}) on the modulation frequency, discussing CW stability from the analytical expressions of $\kappa_{1,2}$ for arbitrary nonzero values of $\Omega$ is far from being an easy task. Therefore we resort to a global analysis, by mapping the two eigenvalues onto a plane Re($\kappa$)-Im($\kappa$) describing a two-dimensional complex parameter space, where Re($\kappa$) and Im($\kappa$) are real and imaginary parts respectively of the eigenvalue $\kappa$. In this parametric representation, the modulation frequency $\Omega$ plays the role of a parameter and so can span a broad range of values, which in our case will be the finite interval $-5\leq\Omega\leq5$. The first figures we consider are parametric representations of Im($\kappa$) as a function of Re($\kappa$), for some selected combinations of values of key characteristic parameters of the model. To be more explicit, the four graphs in figs. \ref{fig1} and \ref{fig2} represent Im($\kappa$) as a function of Re($\kappa$) in the anomalous dispersion regime ($\delta<0$: fig. \ref{fig1}) and normal dispersion regime ($\delta>0$: fig. \ref{fig2}) respectively, for $K=2$, $3$, $4$ and $5$. Values of model parameters are given in the captions, and different curves in each graph correspond to different values of the radiative recombination coefficient $a$. Recall that the sign of $\delta$ determines the dispersion regime \cite{r22,r23,mbieda}, indeed a positive $\delta$ corresponds to a normal group-velocity dispersion, whereas a negative $\delta$ will correspond to an anomalous group-velocity dispersion well known \cite{r22,r23,mbieda} to favor the generation of pulse structures, of course provided the intrinsic refractive index of the host medium is of a self-focusing Kerr nonlinearity.  

  Fig. \ref{fig1} suggests that small values of $K$ are expected to favor laser self-starting in the anomalous dispersion regime. As $K$ increases, the real part of the largest eigenvalue gradually shifts to the negative branch and consequently, the CW regime is stabilized. On the contrary fig. \ref{fig2} indicates that in the normal dispersion regime, CW operation will be favored for small values of the multiphoton absorption rate $K$. As $K$ increases, CW modes become unstable thus favoring laser self-starting.
  
\section{Nonlinear regime}
In the full nonlinear regime, solutions to the laser equation eq. (\ref{j1}) are high-intensity fields which can be represented as real-amplitude pulses, undergoing spatio-temporal modulations i.e. \cite{soto,mbieda}:
\begin{equation}
	u(z,t)=g(z,t)\exp i\left[\phi(z,t)-\omega z \right],\label{u} 
\end{equation}
where $g$ is the real amplitude and $\phi$ is the modulation phase. We introduce a reduced time as $\tau=t-vz$, in which $v$ is the pulse inverse velocity such that $\omega$ in formula (\ref{u}) emerges more explicitly as a nonlinear shift in the propagation constant. Letting $g(z,t)\equiv g(\tau)$ and $\phi(z,t)\equiv \phi(\tau)$, inserting eq. (\ref{u}) into eq. (\ref{j1}) and eq. (\ref{j2}) and separating real parts from imaginary parts, we obtain the following set of coupled first-order nonlinear ordinary differential equations:
\begin{eqnarray}
	\left(vM+\omega+\delta M^2+\gamma\omega_{0}\tau_{0}\rho \right)g -\delta y'+\sigma g^3&=&0, \label{ge} \\
	\left(\delta M'-\gamma\rho \right)g+\left( v+2\delta M\right) y-\mu g^{2K-1} &=&0, \label{gb} \\
	\rho'-\nu g^2 \rho -\alpha g^{2K}+a\rho^2&=&0, \label{m2}
\end{eqnarray}
with $y=g'$, $M=\phi'$ and the prime symbol refers to derivative with respect to $\tau$. Let us focus on the system dynamics in the particular case $v=0$ \cite{r24,soto,mbieda}. For this value of $v$ the set of coupled first-order nonlinear ordinary differential equations (\ref{ge})-(\ref{m2}) reduces to:
\begin{eqnarray}
M'&=&\frac{\gamma\rho}{\delta}-\frac{2My}{g}+\frac{\mu g^{2K-2}}{\delta}, \nonumber \\
y'&=&\frac{\left( \omega+\delta M^2+\gamma\omega_{0}\tau_{0}\rho\right)g }{\delta}+\frac{\sigma g^3}{\delta}, \nonumber \\
g'&=&y, \nonumber \\
\rho'&=& \nu g^2 \rho +\alpha g^{2K}-a\rho^2. \label{m3}
\end{eqnarray}
 Our first interest will be on the singular solutions to this system, which are their fixed points, with the aim to probe the effects of important characteristic parameters of the model such as the radiative recombination coefficient $a$ and the multiphoton absorption rate $K$, on equilibrium solutions of the laser amplitude $g$ and instantaneous frequency $M$, as well as of the electron plasma density $\rho$.  
\subsection{Fixed points}
Singular solutions to the set of first-order nonlinear ordinary differential equations (\ref{m3}), which are their fixed points, are the roots of the following nonlinear system:
\begin{eqnarray}
		\frac{\gamma\rho}{\delta}+\frac{\mu g^{2K-2}}{\delta}&=&0, \\
		\frac{\left( \omega+\delta M^2+\gamma\omega_{0}\tau_{0}\rho\right)g }{\delta}+\frac{\sigma g^3}{\delta}&=&0, \\
		y&=&0, \\
		\nu g^2\rho +\alpha g^{2K}-a\rho^2&=&0. \label{2a}
\end{eqnarray}
From this system we derive:
\begin{equation}
M^2=\frac{1}{\delta}\left[\mu\omega_{0}\tau_{0}g^{2K-2}-\omega-\sigma g^2 \right], \qquad \rho=g^{K}\,\sqrt{\left(\frac{\alpha\gamma-\mu\nu}{a\gamma} \right)}. \label{n1}
\end{equation}
Remarkable enough formula (\ref{n1}) suggests that irrespective of the value of $K$, the electron plasma density $\rho$ will be zero when the laser amplitude $g$ is zero. \par
\begin{figure}\centering
\begin{minipage}[c]{0.5\textwidth}
\includegraphics[width=3.in, height= 2.5in]{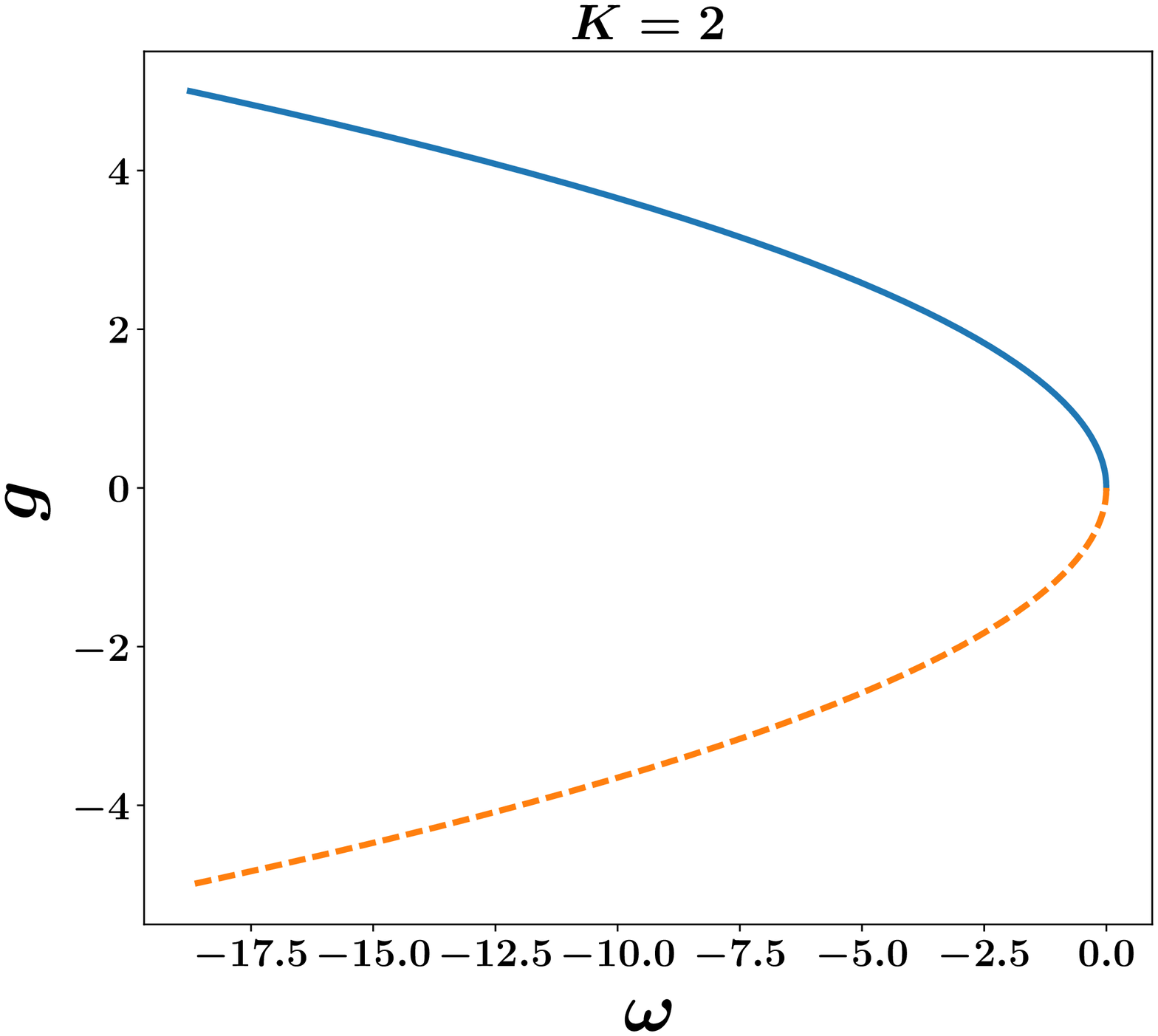}
\end{minipage}%
\begin{minipage}[c]{0.5\textwidth}
\includegraphics[width=3.in, height= 2.5in]{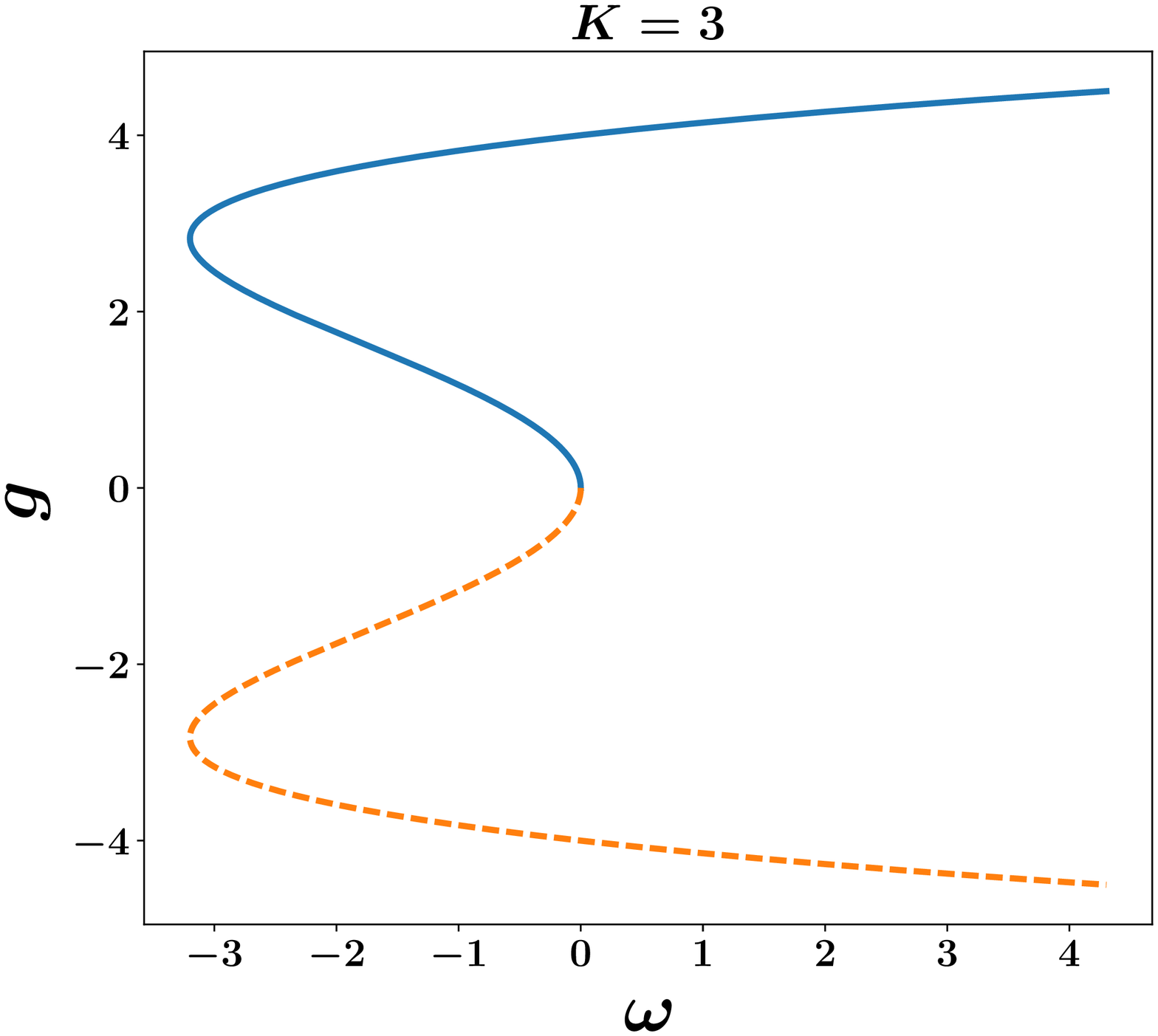}
\end{minipage}\\
\vspace{0.5truecm}
\begin{minipage}[c]{0.5\textwidth}
\includegraphics[width=3.in, height= 2.5in]{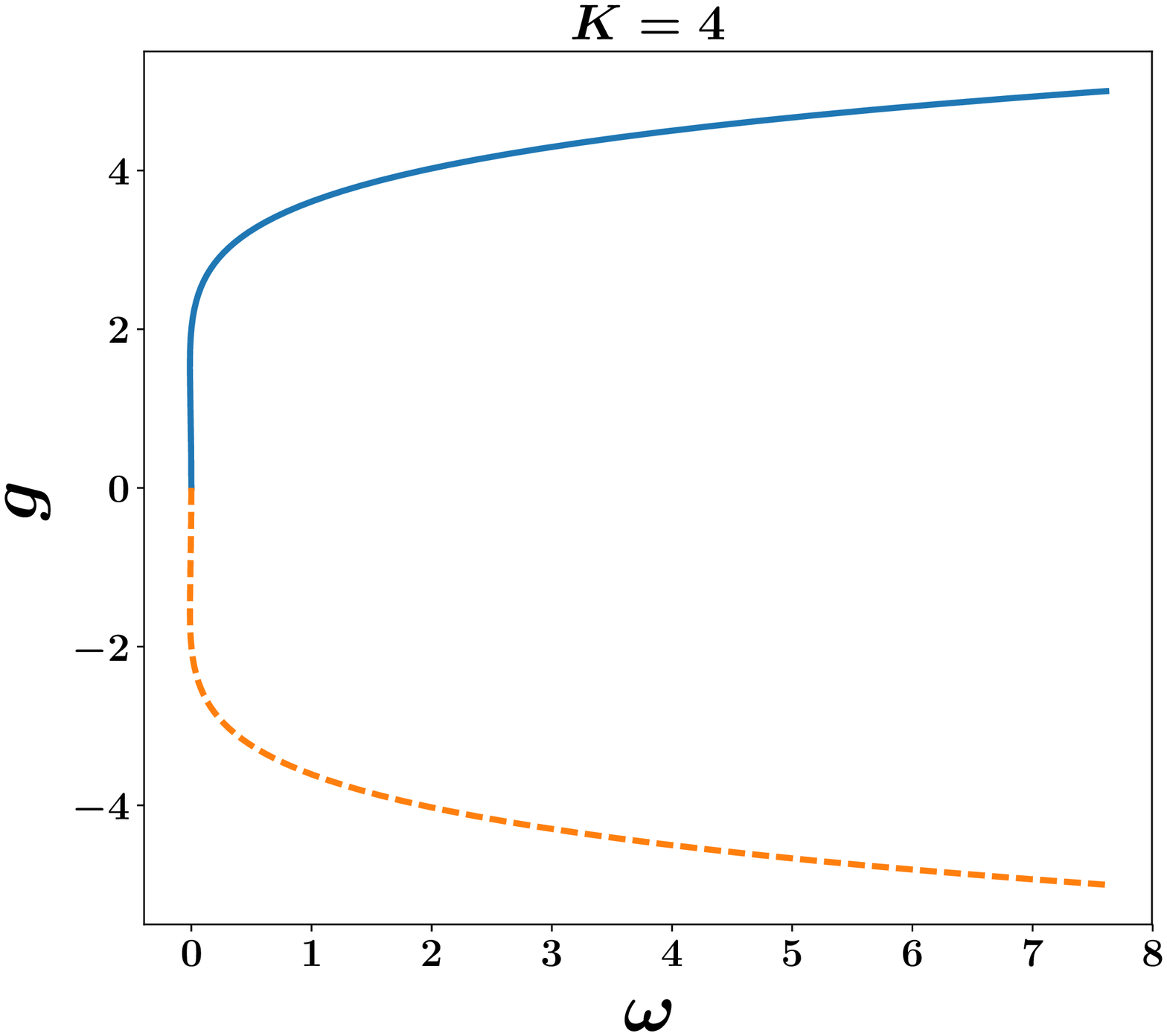}
\end{minipage}%
\begin{minipage}[c]{0.5\textwidth}
\includegraphics[width=3.in, height= 2.5in]{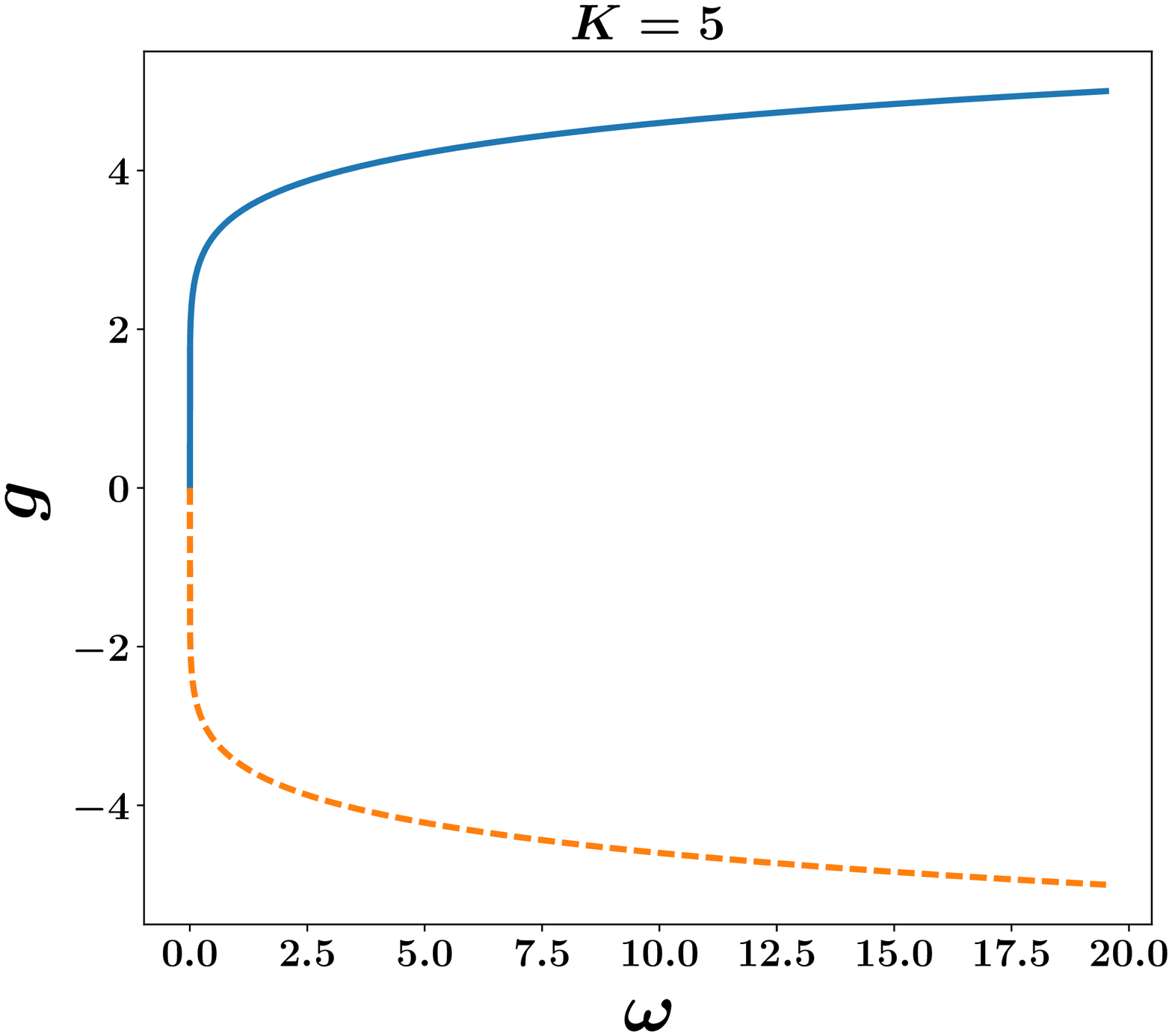}
\end{minipage}
\caption{(Color online) Fixed points of the laser amplitude $g$ as a function of $\omega$, for $K=2, 3, 4, 5$, $\sigma=0.8$, $\omega_0\tau_0=0.2$, $\mu=0.25$.}{\label{fig3}}
  \end{figure}

\begin{figure}\centering
\begin{minipage}[c]{0.5\textwidth}
\includegraphics[width=3.in, height= 2.5in]{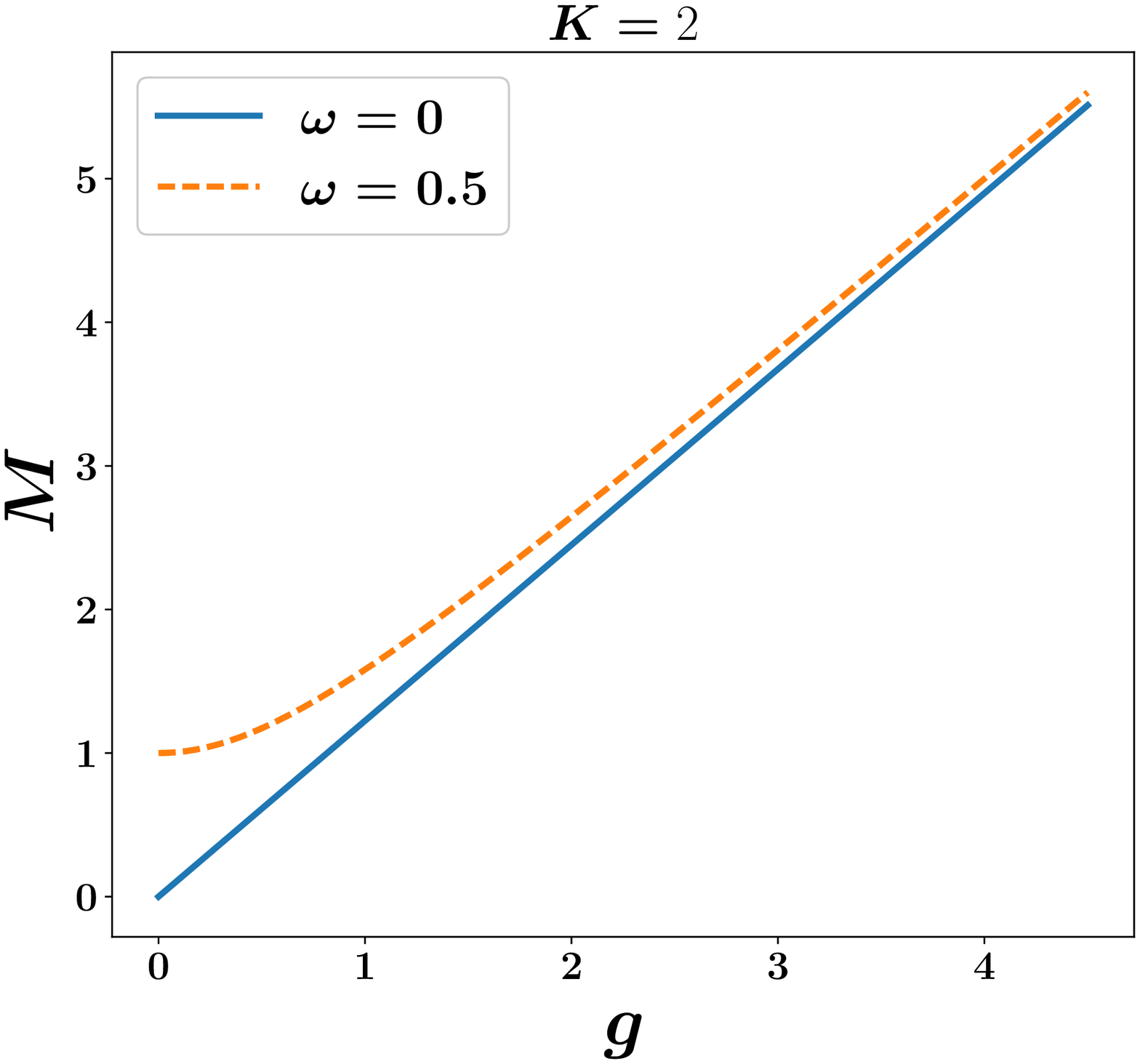}
\end{minipage}%
\begin{minipage}[c]{0.5\textwidth}
\includegraphics[width=3.in, height= 2.5in]{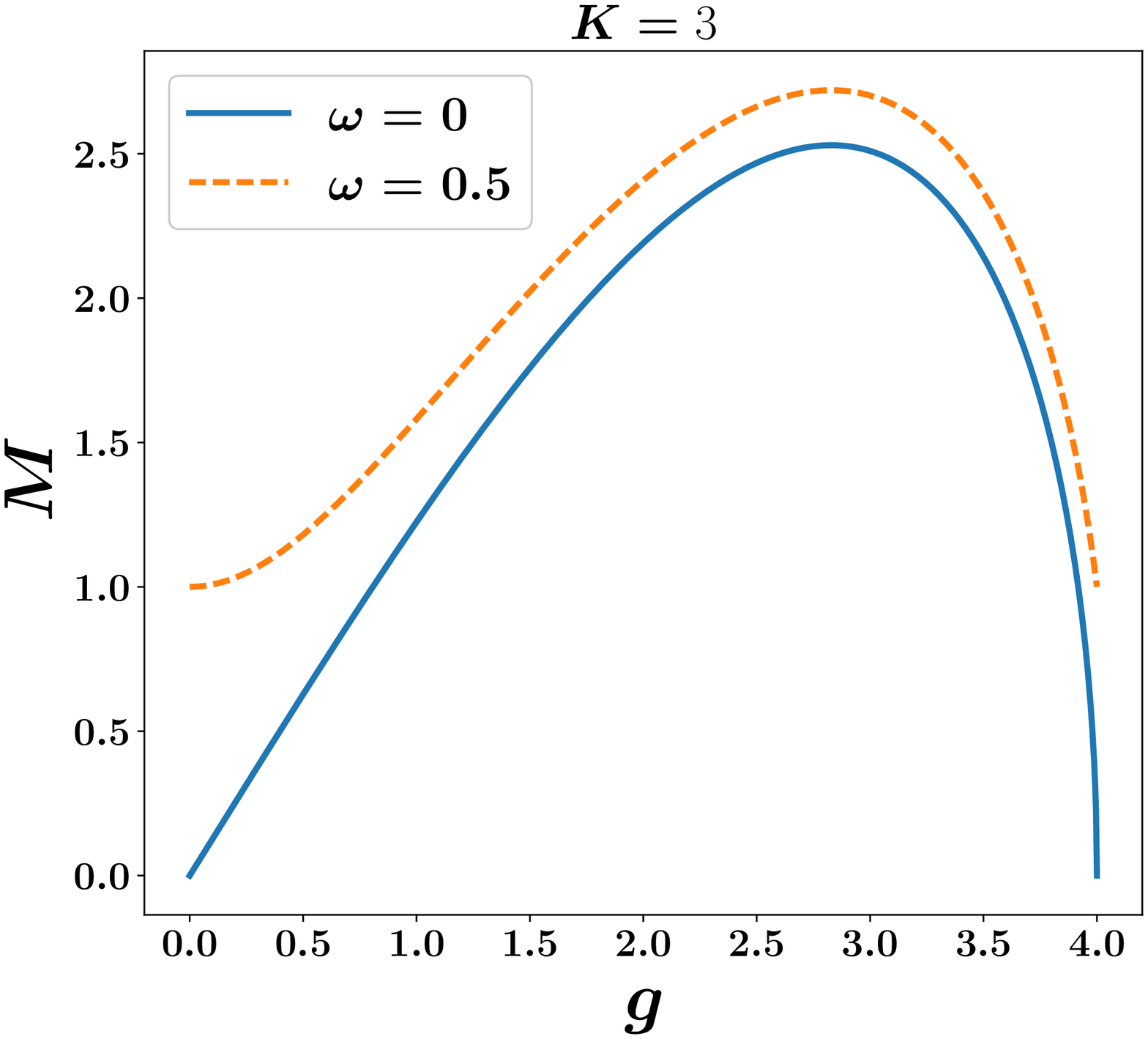}
\end{minipage}\\
\vspace{0.5truecm}
\begin{minipage}[c]{0.5\textwidth}
\includegraphics[width=3.in, height= 2.5in]{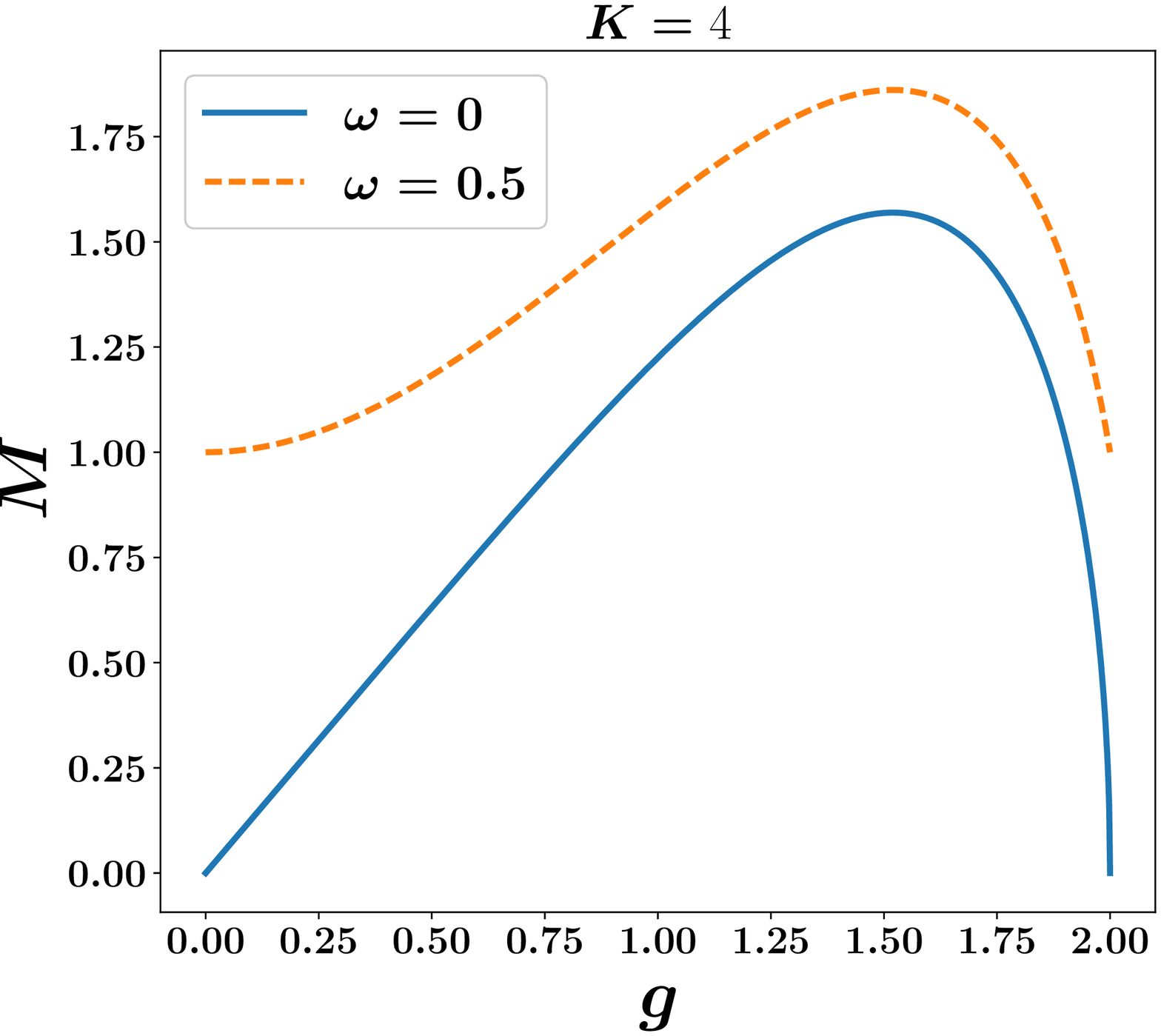}
\end{minipage}%
\begin{minipage}[c]{0.5\textwidth}
\includegraphics[width=3.in, height= 2.5in]{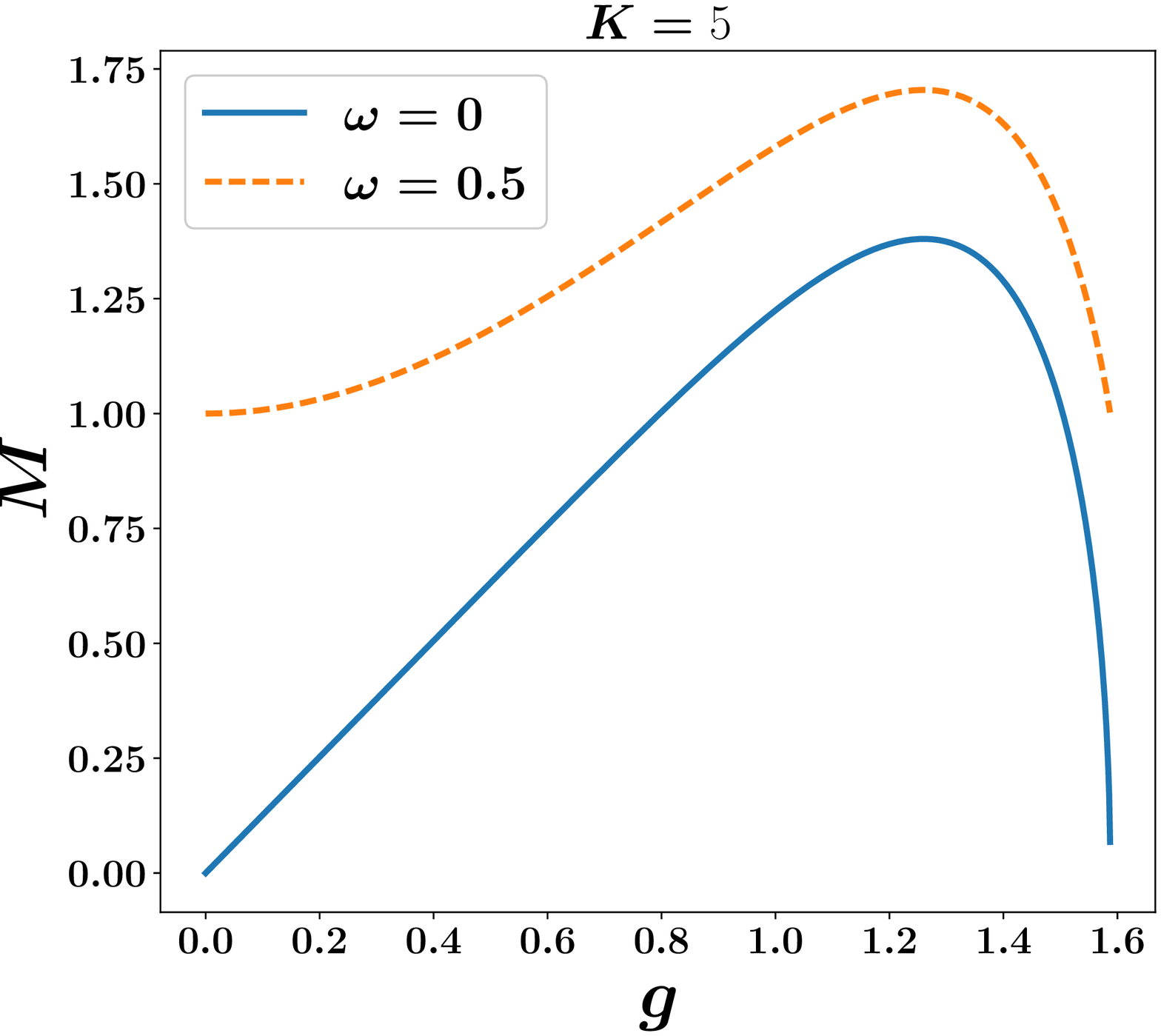}
\end{minipage}
\caption{(Color online) Variation of the instantaneous frequency $M$ with the amplitude $g$ of laser, for $K=2, 3, 4, 5$ and two different values of $\omega$ (Solid curve correspond to $\omega=0$, and dashed curve to $\omega=0.5$). Values of other parameters are $\sigma=0.8$, $\omega_0\tau_0=0.2$, $\delta=-0.5$ , $\mu=0.25$.}{\label{fig4}}
  \end{figure}

\begin{figure}\centering
\begin{minipage}[c]{0.5\textwidth}
\includegraphics[width=3.in, height= 2.5in]{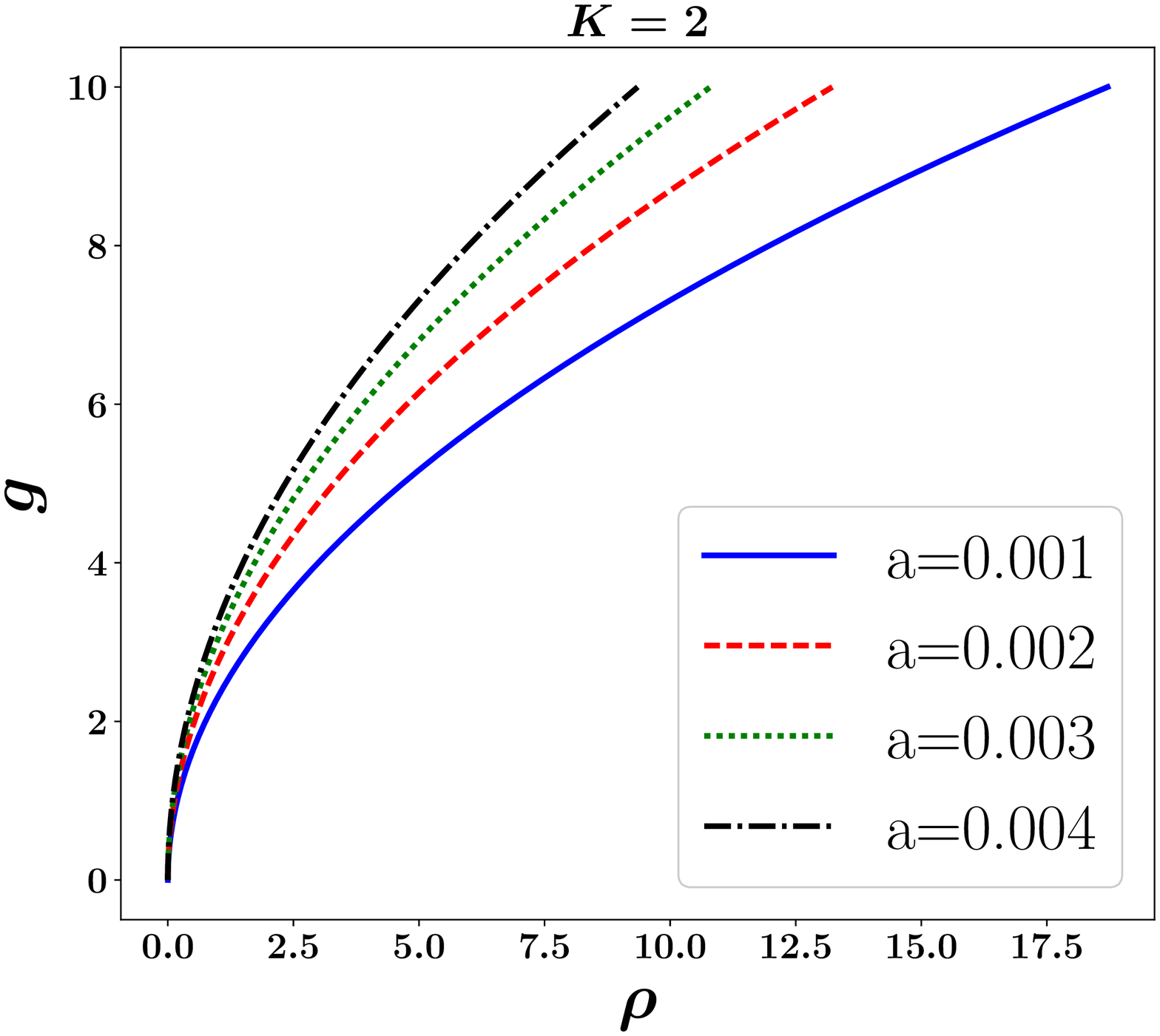}
\end{minipage}%
\begin{minipage}[c]{0.5\textwidth}
\includegraphics[width=3.in, height= 2.5in]{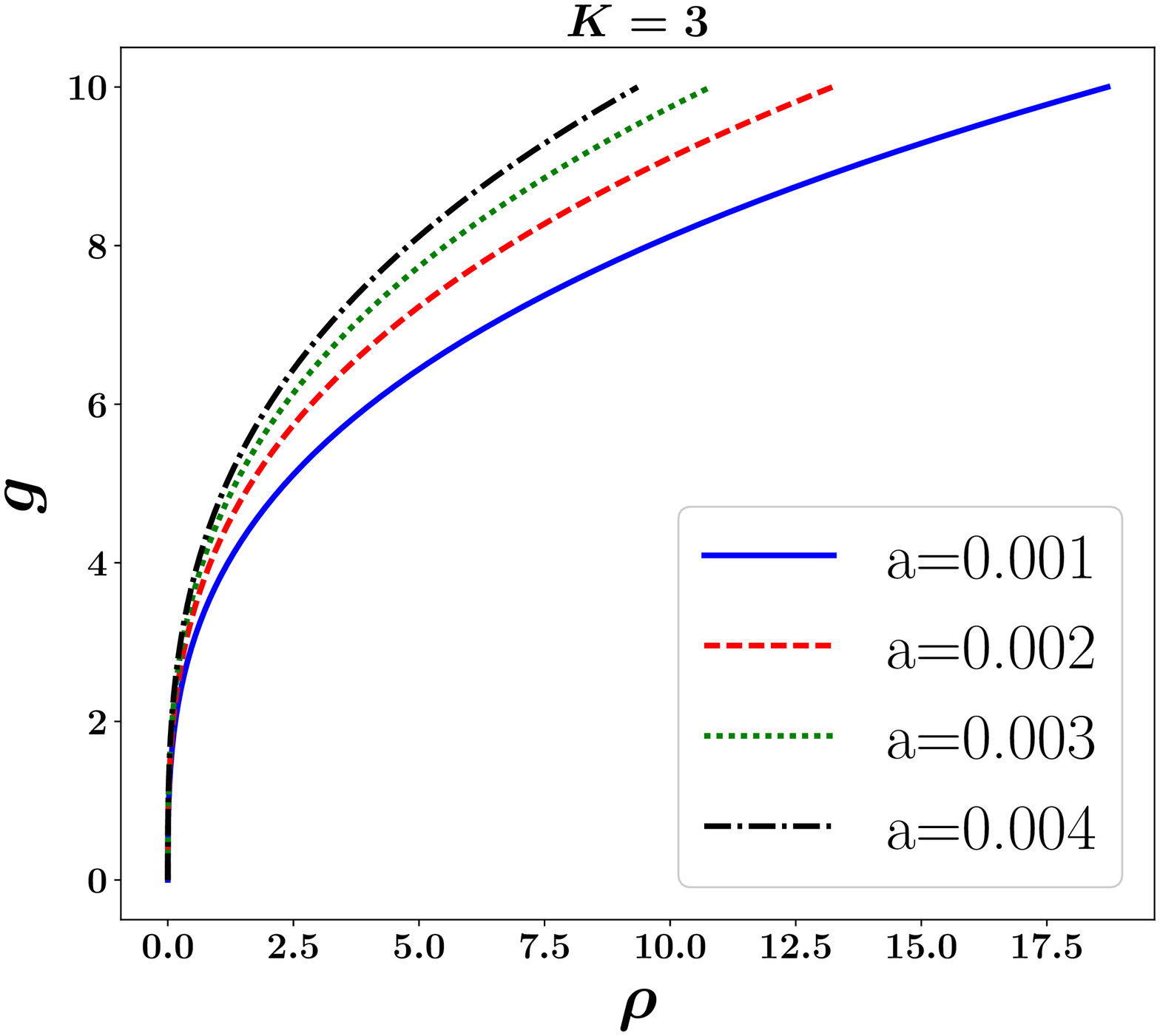}
\end{minipage}\\
\vspace{0.5truecm}
\begin{minipage}[c]{0.5\textwidth}
\includegraphics[width=3.in, height= 2.5in]{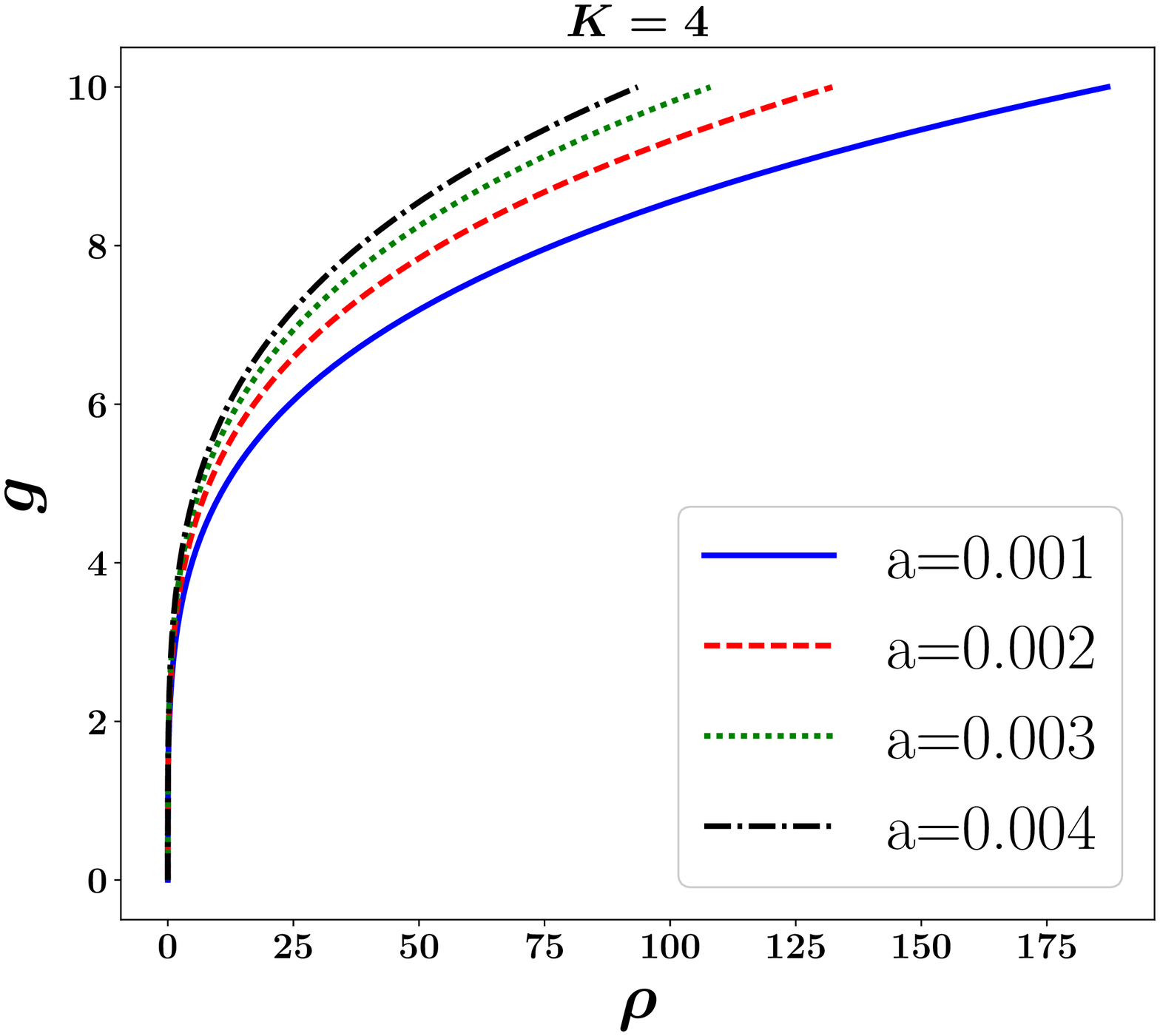}
\end{minipage}%
\begin{minipage}[c]{0.5\textwidth}
\includegraphics[width=3.in, height= 2.5in]{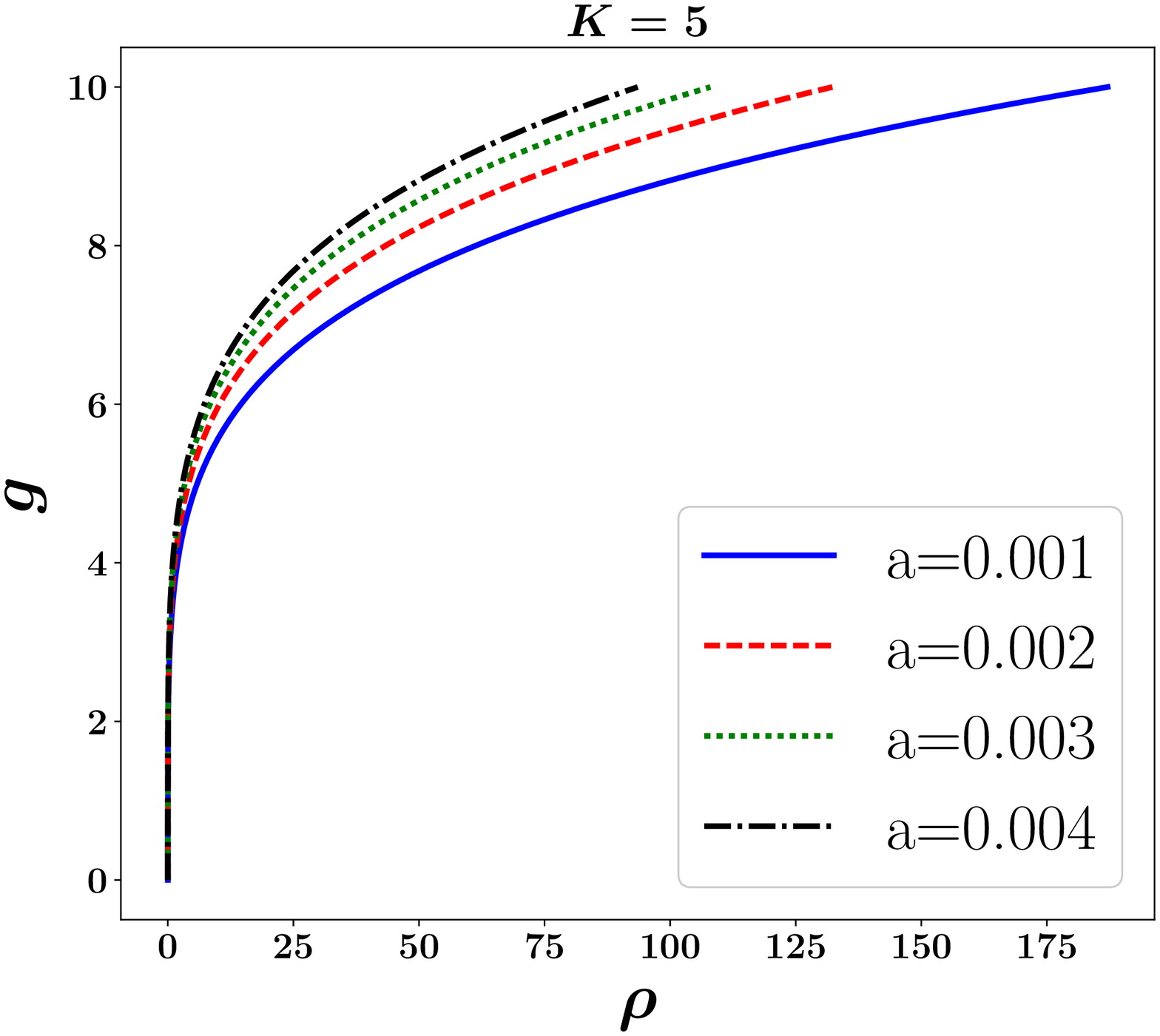}
\end{minipage}
\caption{(Color online) Variation of the laser amplitude $g$ with the electron plasma density $\rho$, for $K=2, 3, 4, 5$ and different values of the radiative recombination coefficient $a$ indicated in the graphs. $\nu=0.1$ $\alpha=0.6$, $\gamma=0.1$ , $\mu=0.25$.}{\label{fig5}}
  \end{figure}

 
Fig. \ref{fig3} represents the variations of the two fixed points of the laser amplitude $g$ as a function of $\omega$, for $K=2, 3, 4, 5$ and values of model parameters given in the captions. It should be noted that by fixed points of $g$ we understand its extrema, i.e. its maximum and minimum which are obtained by annihilating $M$ in formula (\ref{n1}).   
  According to fig. \ref{fig3}, for $K=2$ the laser dynamics is dominated by weakly nonlinear pulse trains with maximum amplitudes for $\omega=0$. An increase of $K$ (see graphs for $K=3$, $4$ and $5$) favors strongly nonlinear pulse trains of larger amplitudes \cite{soto,mbieda}.   
  We have also plotted the instantaneous frequency $M$ as a function of the amplitude $g$ of laser (fig. \ref{fig4}), and the laser amplitude $g$ as function of the electron plasma density $\rho$ (fig. \ref{fig5}), for four different values of $K$. Remark that the expression of $M$ given in formula (\ref{n1}) does not contains the radiative recombination coefficient $a$ but is controlled mainly by the laser propagation constant $\omega$, whereas $g$ as a function of $\rho$ extracted from eq. (\ref{n1}) depends on $a$ but not on $\omega$. Therefore, we have chosen to plot $M$ as a function of $g$ by considering two cases i.e., the case $\omega=0$ and the case of finite nonzero value of $\omega$ as one sees in fig. \ref{fig4}. 
  The different curves in the graphs of fig. \ref{fig4} show that $M$ is enhanced by an increase of $g$, and that there is a threshold value of the amplitude beyond which the instantaneous frequency is expected to decrease to zero. As it is apparent, this threshold value of $g$ (and consequently of $M$) is decreased with an increase of $K$.
  In fig. \ref{fig5}, the fixed point of the electron plasma density is always zero at zero amplitude of the laser whatever the value of $K$, consistently with what we learned from formula (\ref{n1}). However there is a drastic increase of $\rho$ with an increase of $K$, meaning that for the same laser intensity the stronger the multiphoton absorption processes the larger will be the electron plasma density. 
  
\subsection{Pulse structures}
Considering the full nonlinear dynamics of the system, the set of first-order ordinary differential equations (\ref{m3}) was solved numerically using a sixth-order Runge-Kutta algorithm adapted from ref.. \cite{r30}. Because eqs. (\ref{j1})-(\ref{j2}) involve several parameters, all of which cannot be varied in this study, we fixed most parameters except two ones i.e. the multiphoton absorption rate $K$, which was given the four different values $K=2, 3, 4, 5$, and the radiative recombination coefficient $a$ which was varied in three distinct ranges of values where three distinct behaviors were noticed. 
\par
\begin{figure}\centering
\begin{minipage}[c]{0.5\textwidth}
\includegraphics[width=3.in, height= 2.5in]{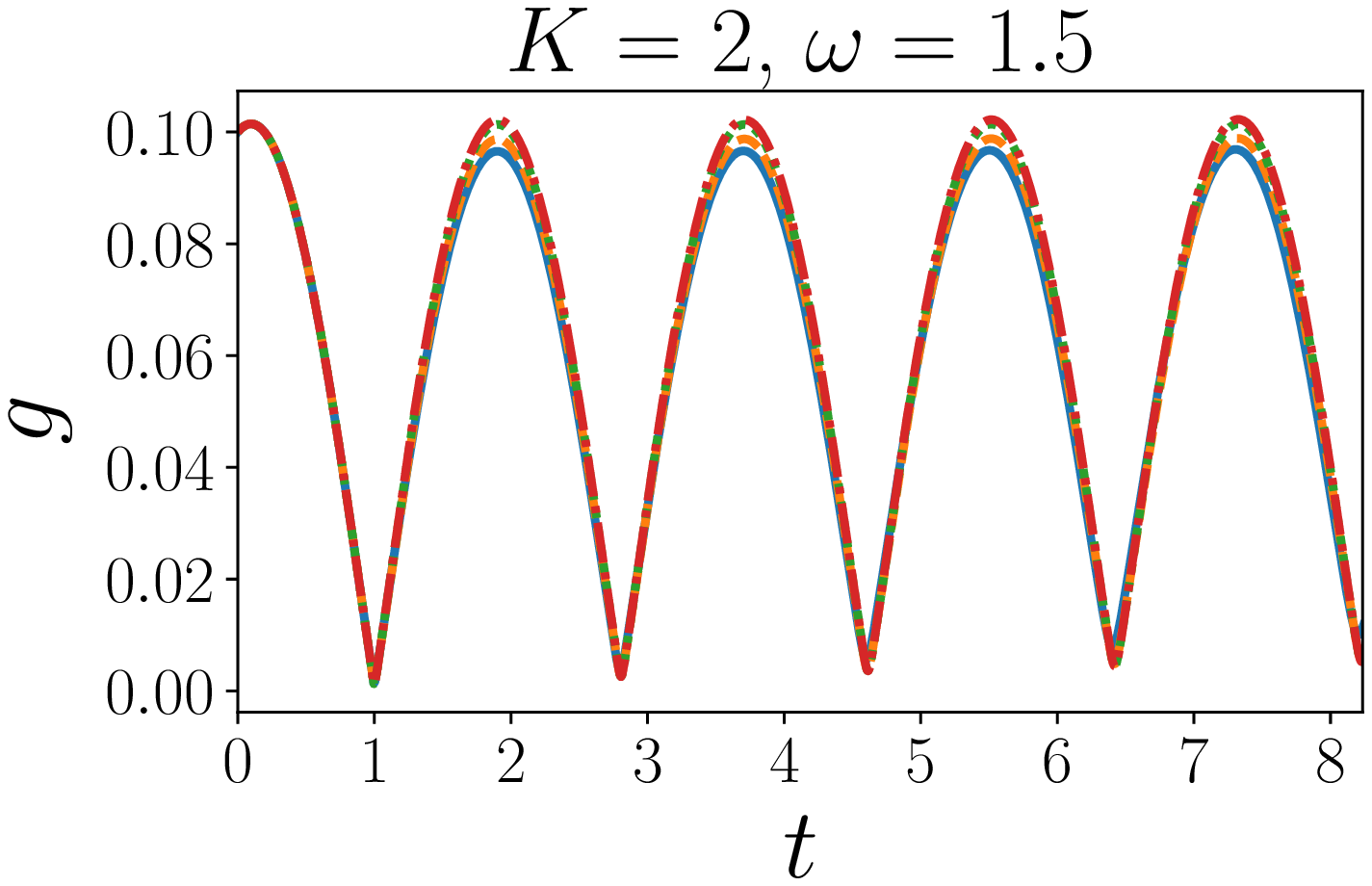}
\end{minipage}%
\begin{minipage}[c]{0.5\textwidth}
\includegraphics[width=3.in, height= 2.5in]{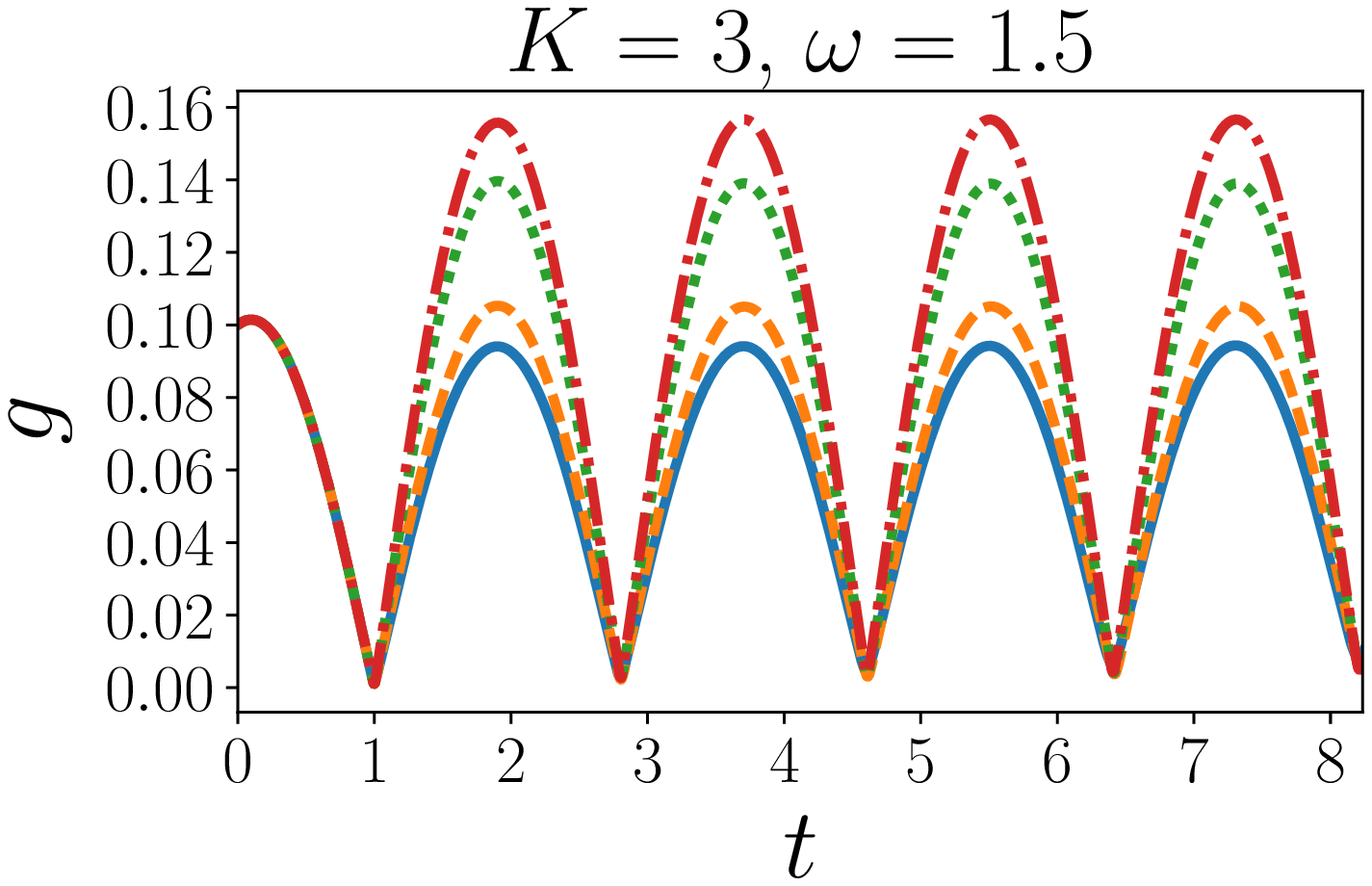}
\end{minipage}\\
\vspace{0.5truecm}
\begin{minipage}[c]{0.5\textwidth}
\includegraphics[width=3.in, height= 2.5in]{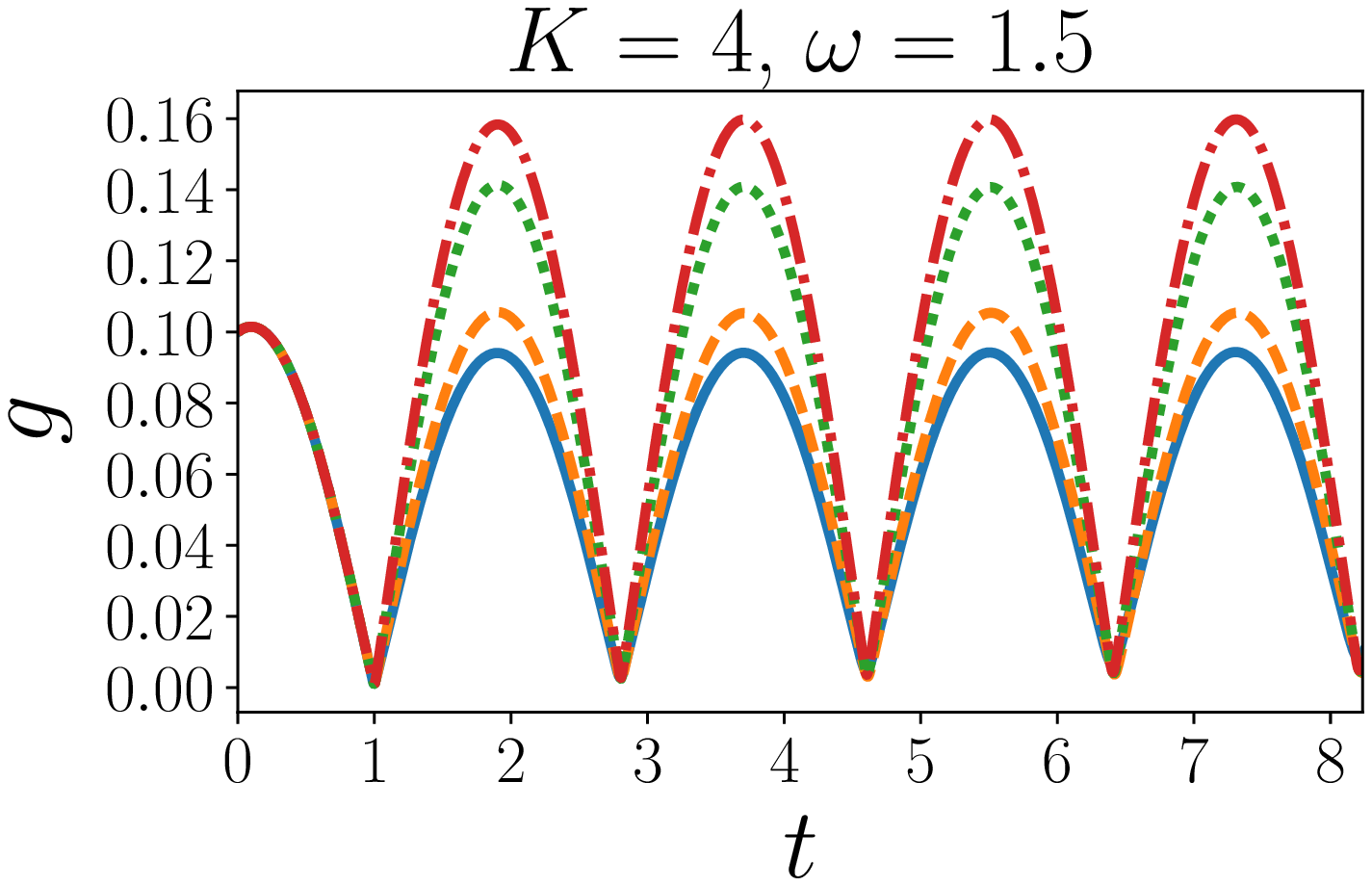}
\end{minipage}%
\begin{minipage}[c]{0.5\textwidth}
\includegraphics[width=3.in, height= 2.5in]{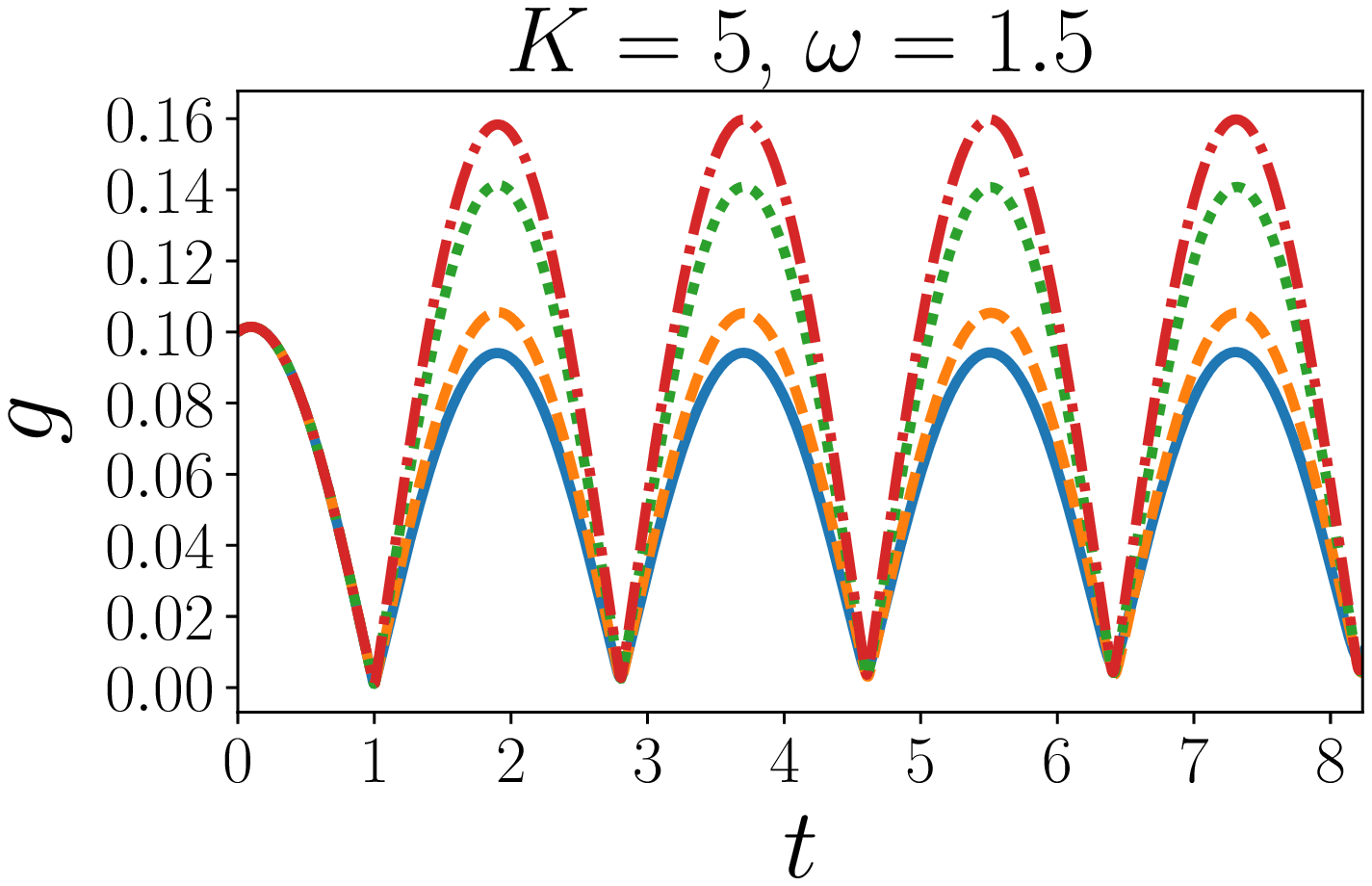}
\end{minipage}
   \caption{(Color online) Time variation of the laser amplitude $g$, for $K=2, 3, 4, 5$ and different values of the radiative recombination coefficient $a$. $\nu=0.1$ $\alpha=0.6$, $\gamma=0.1$ , $\mu=0.25$. Values of $a$ are, from the smallest to the largest amplitudes: 0, 3.9, 5.6, 6.}{\label{fig6}}
  \end{figure}

\begin{figure}\centering
\begin{minipage}[c]{0.5\textwidth}
\includegraphics[width=3.in, height= 2.5in]{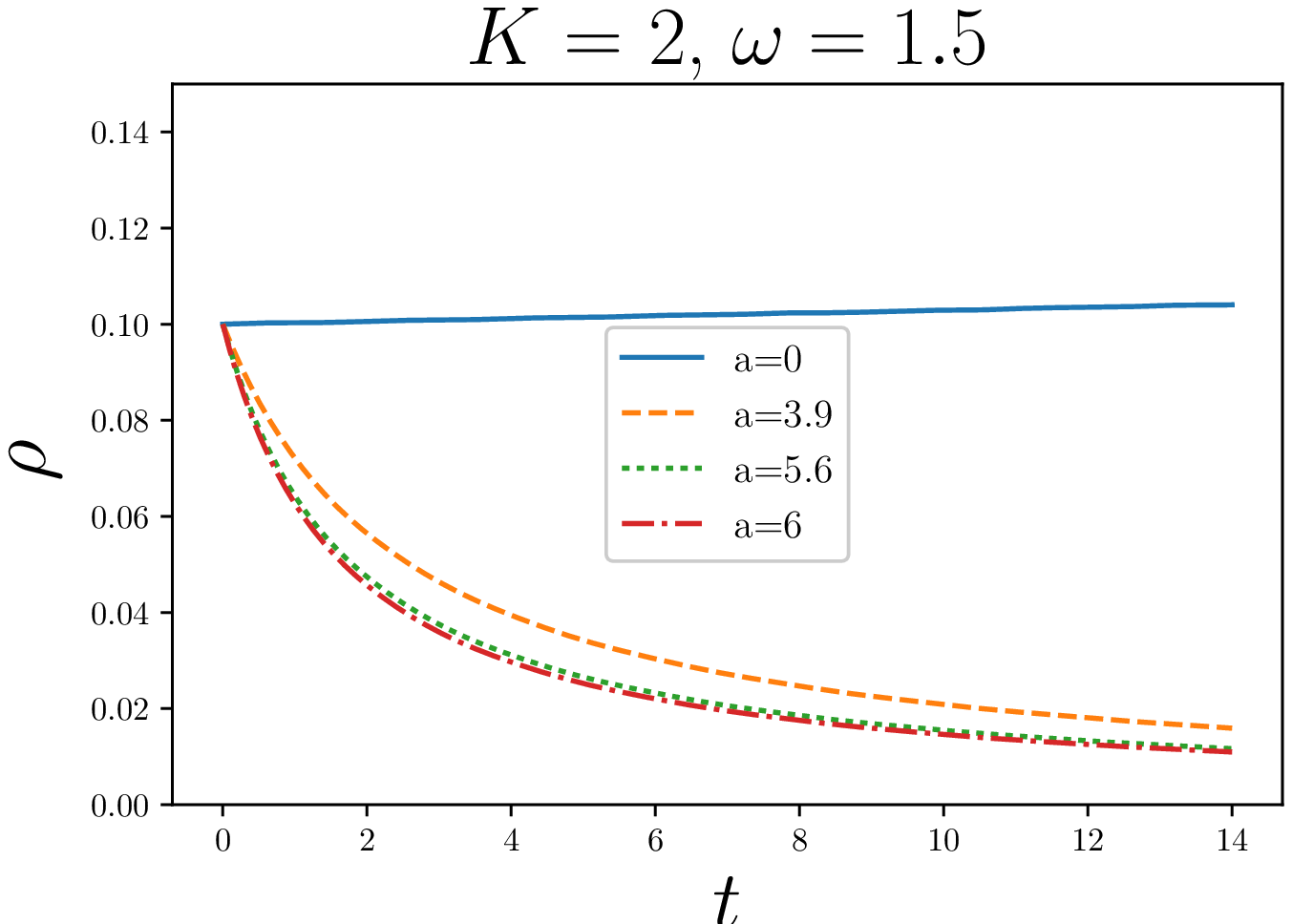}
\end{minipage}%
\begin{minipage}[c]{0.5\textwidth}
\includegraphics[width=3.in, height= 2.5in]{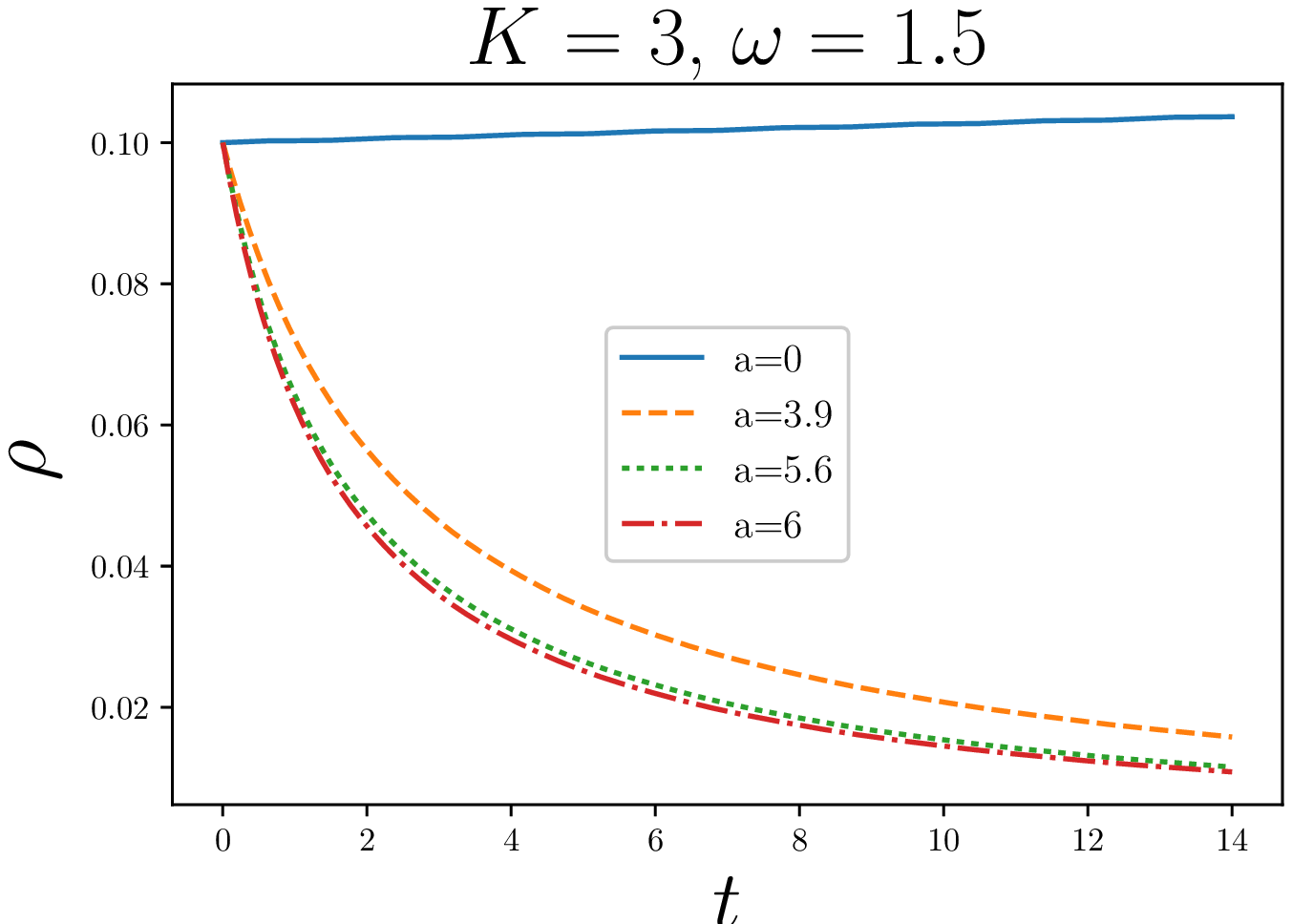}
\end{minipage}\\
\vspace{0.5truecm}
\begin{minipage}[c]{0.5\textwidth}
\includegraphics[width=3.in, height= 2.5in]{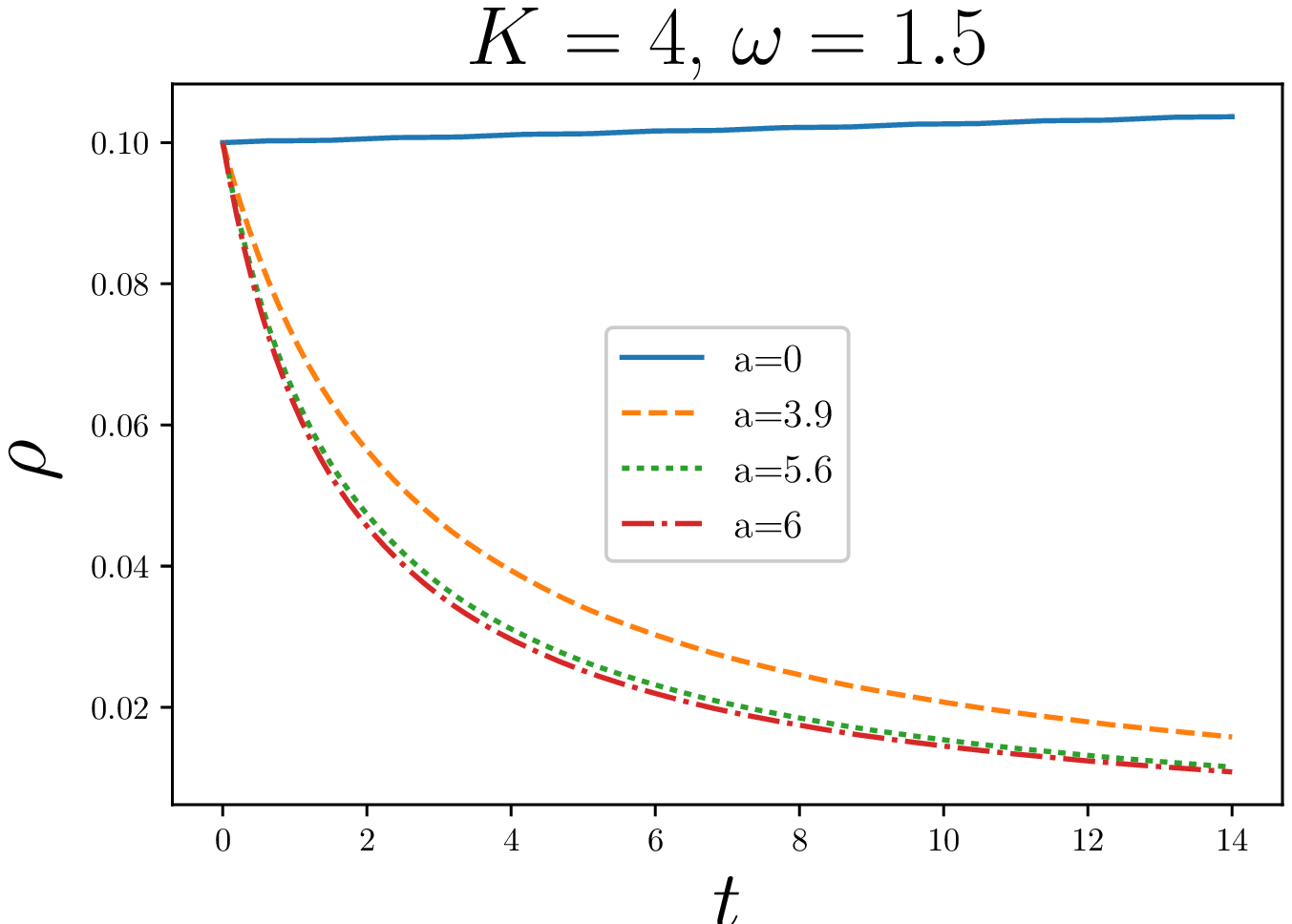}
\end{minipage}%
\begin{minipage}[c]{0.5\textwidth}
\includegraphics[width=3.in, height= 2.5in]{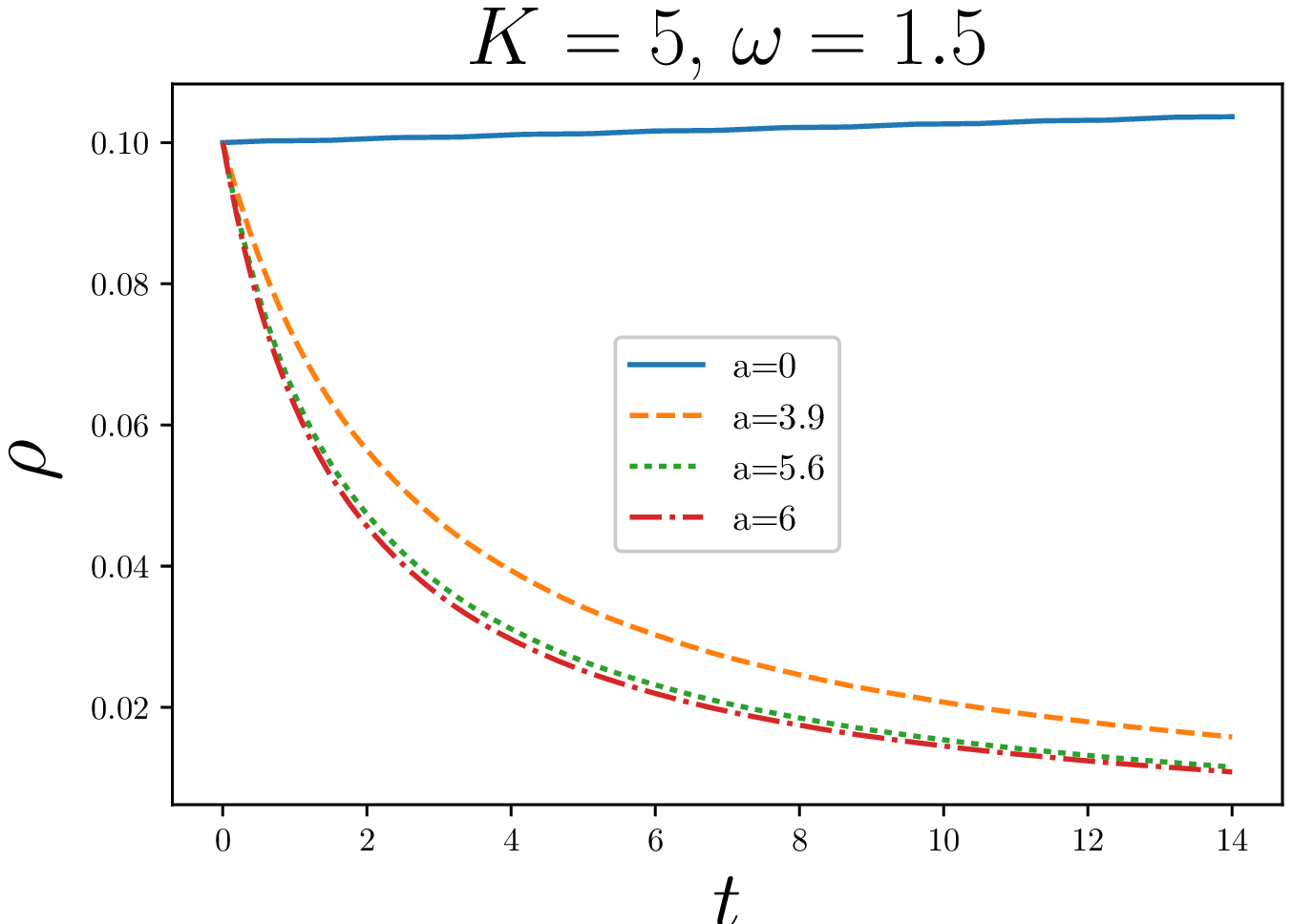}
\end{minipage}
\caption{(Color online) Time variation of the electron plasma density $\rho$, for $K=2, 3, 4, 5$ and values of $a$ indicated in the graphs. $\sigma=0.8$, $\omega_0\tau_0=0.8$, $\delta=-0.5$ , $\mu=0.5$, $\nu=0.5$, $\gamma=0.18$, $\alpha=0.6$.}{\label{fig8}}
  \end{figure}

 Figs. \ref{fig6} and \ref{fig8} are time variations of the laser amplitude $g(t)$ and of the electron plasma density $\rho(t)$, for four distinct values of $K$ and values of model parameters listed in the figure captions.   
Note that for each graph, we plotted dynamical quantities for different values of $a$ starting with $a=0$ (i.e. when there is no radiative recombination). Our objective in so doing was to highlight the qualitative and quantitative influences of $a$, on the system dynamics. \par
In fig. \ref{fig6}, the laser amplitude $g(t)$ is manifestly a pulse train with a maximum varying only weakly with $a$ for $K=2$. However, as $K$ is increased the maximum of $g$ gets more and more large. Note the fall-off of $g(t)$ with time, which is more and more pronounced as $K$ gets larger and for $a=0$. We attribute this fall-off to a damping effect induced by the density of electron plasma i.e. $\rho(t)$, which in this context acts like a laser gain/loss. 
\par
\begin{figure}\centering
\begin{minipage}[c]{0.5\textwidth}
\includegraphics[width=3.in, height= 2.5in]{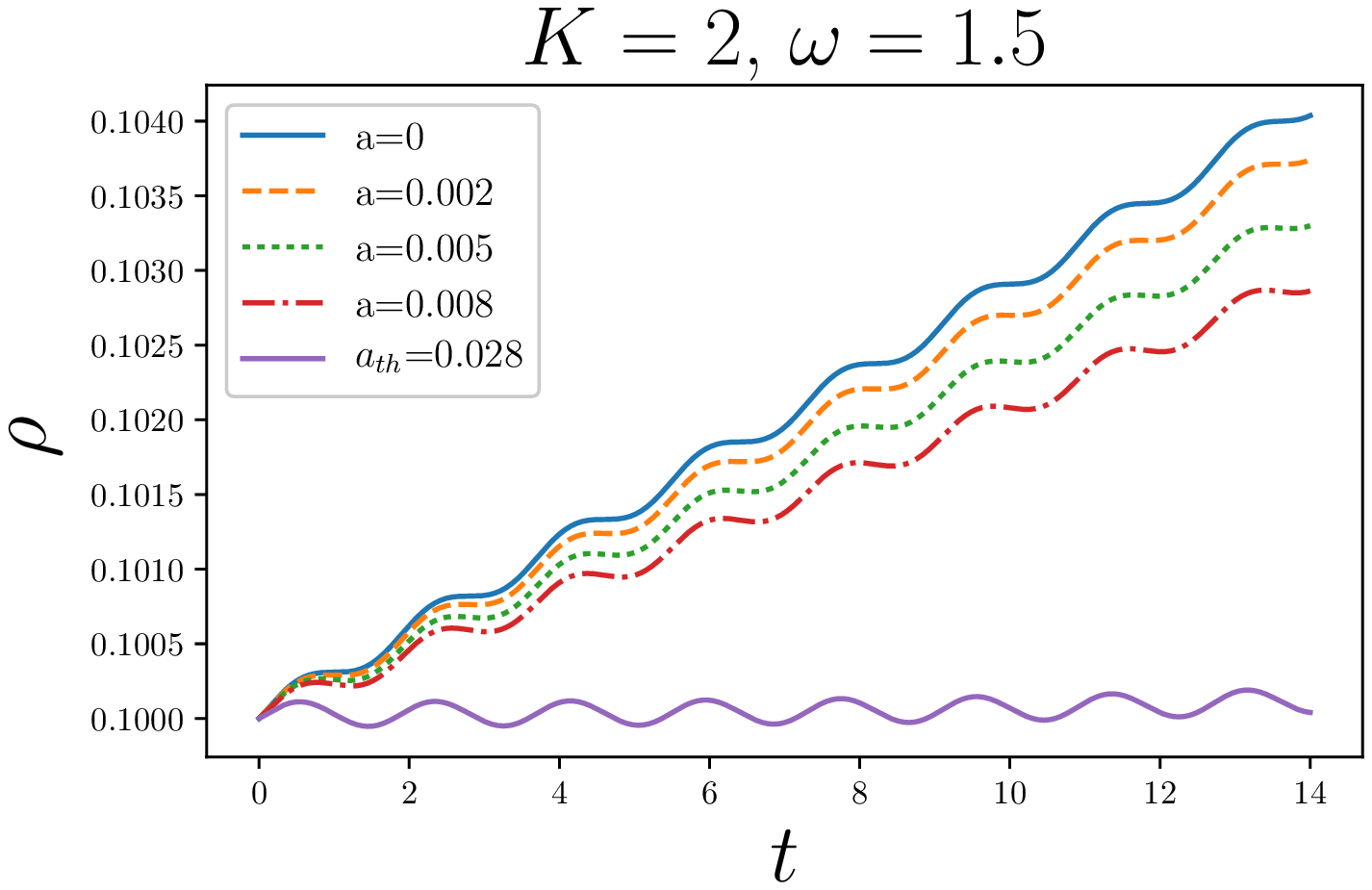}
\end{minipage}%
\begin{minipage}[c]{0.5\textwidth}
\includegraphics[width=3.in, height= 2.5in]{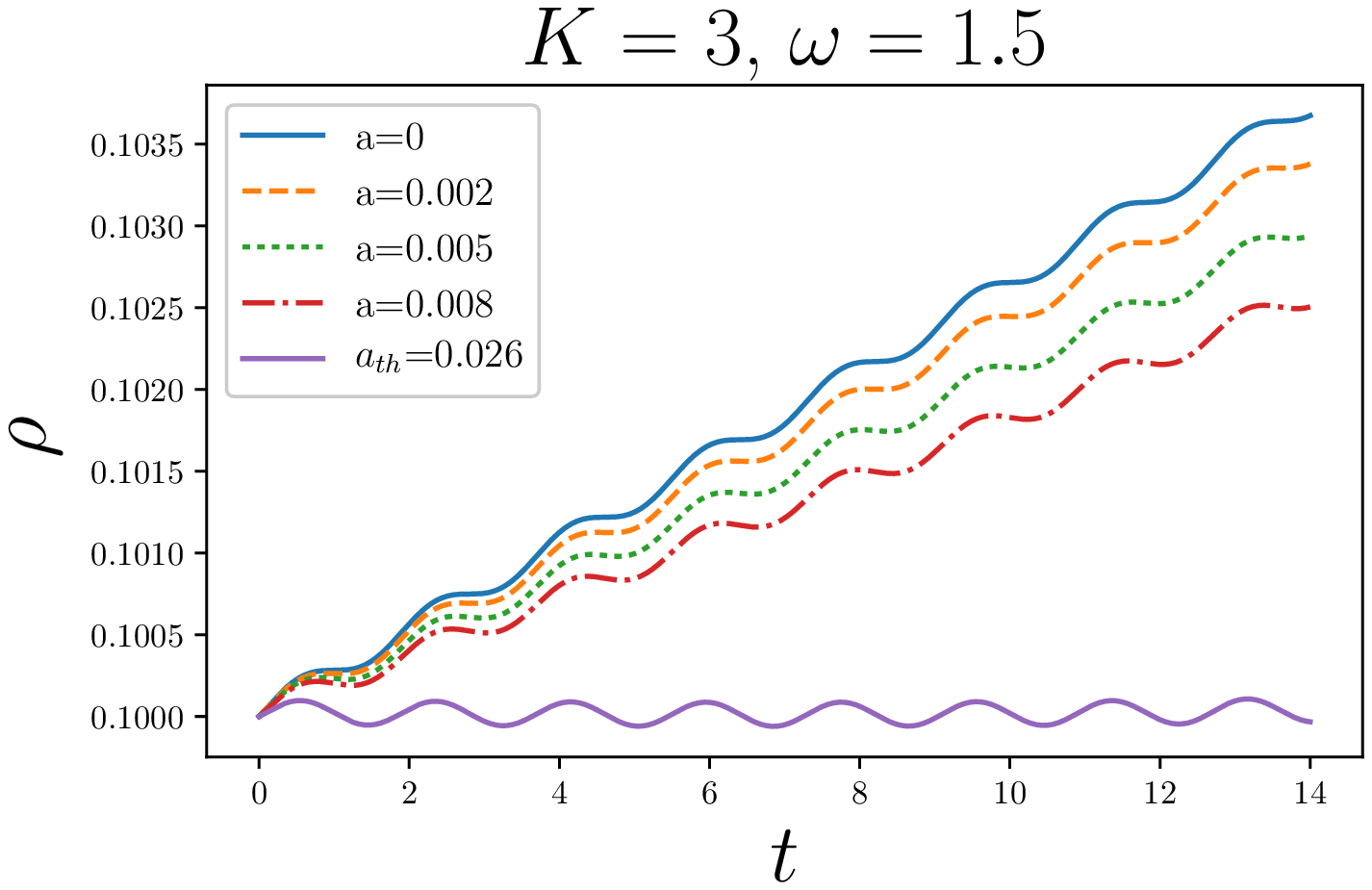}
\end{minipage}\\
\vspace{0.5truecm}
\begin{minipage}[c]{0.5\textwidth}
\includegraphics[width=3.in, height= 2.5in]{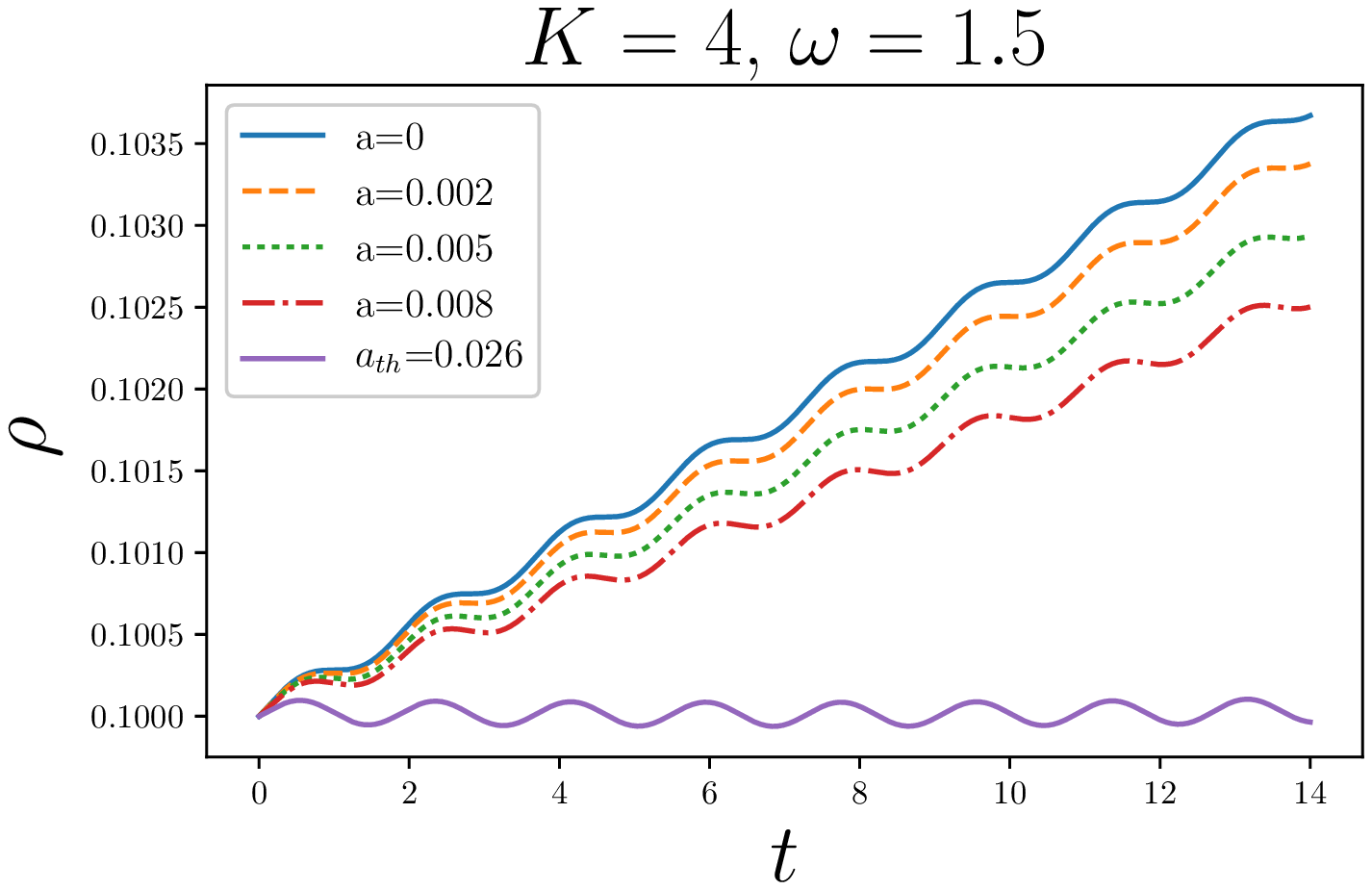}
\end{minipage}%
\begin{minipage}[c]{0.5\textwidth}
\includegraphics[width=3.in, height= 2.5in]{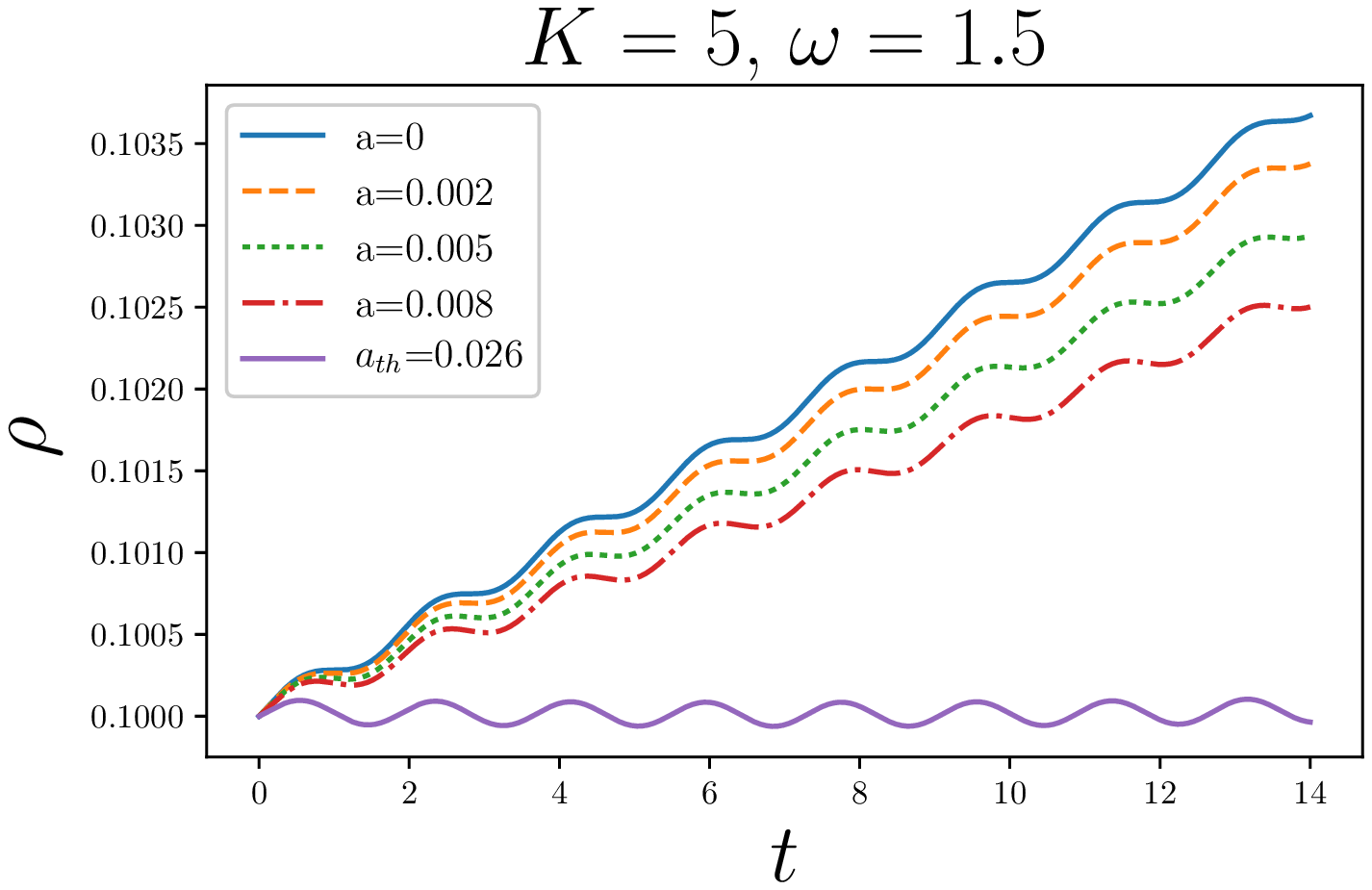}
\end{minipage} 
    \caption{(Color online) Time variation of the electron plasma density $\rho$, for $K=2, 3, 4, 5$ and values of $a$ indicated in the graphs. $\sigma=0.8$, $\omega_0\tau_0=0.8$, $\delta=-0.5$ , $\mu=0.5$, $\nu=0.5$, $\gamma=0.18$, $\alpha=0.6$ .}{\label{fig9}}
  \end{figure}

\begin{figure}\centering
\begin{minipage}[c]{0.5\textwidth}
\includegraphics[width=3.in, height= 2.5in]{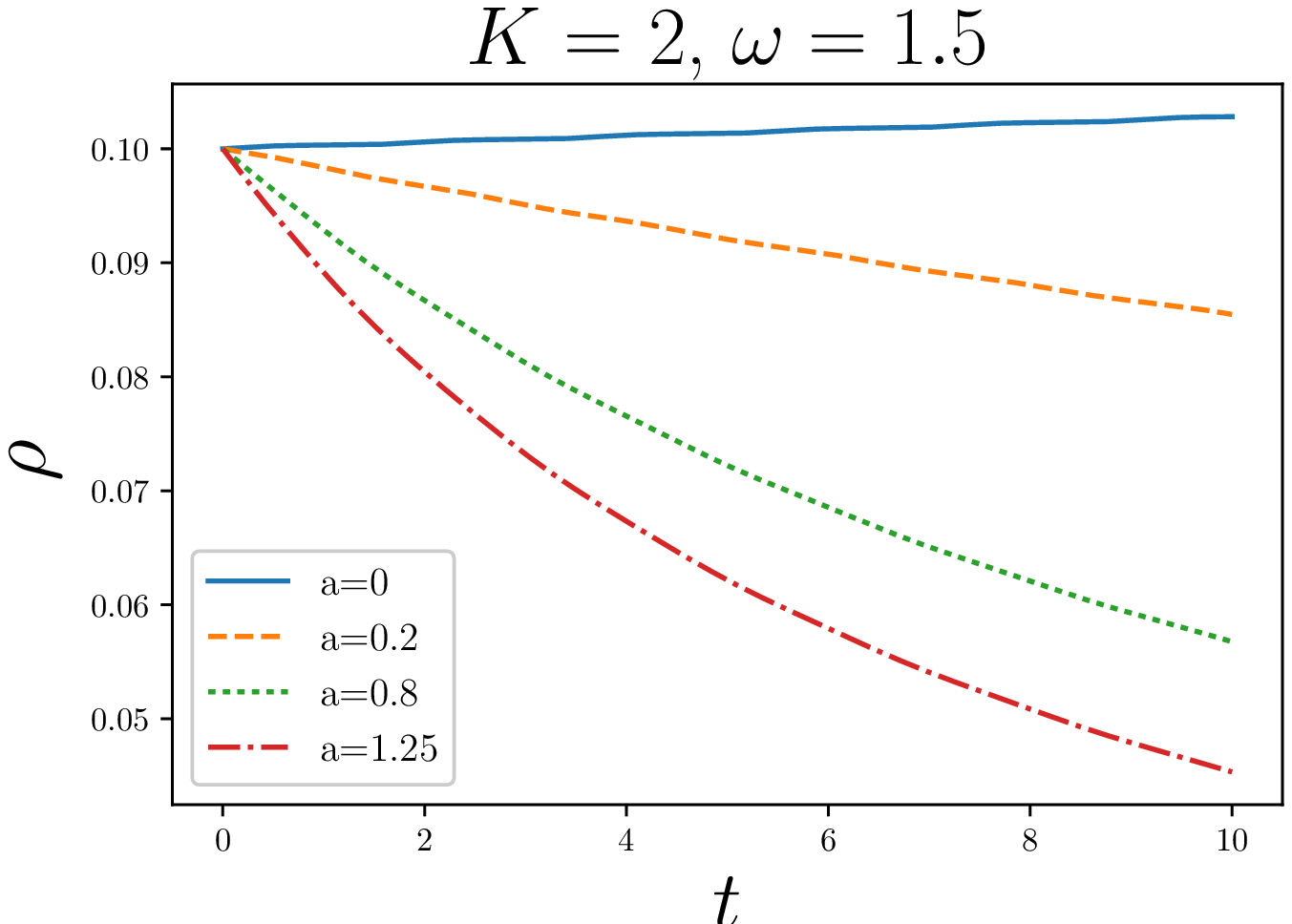}
\end{minipage}%
\begin{minipage}[c]{0.5\textwidth}
\includegraphics[width=3.in, height= 2.5in]{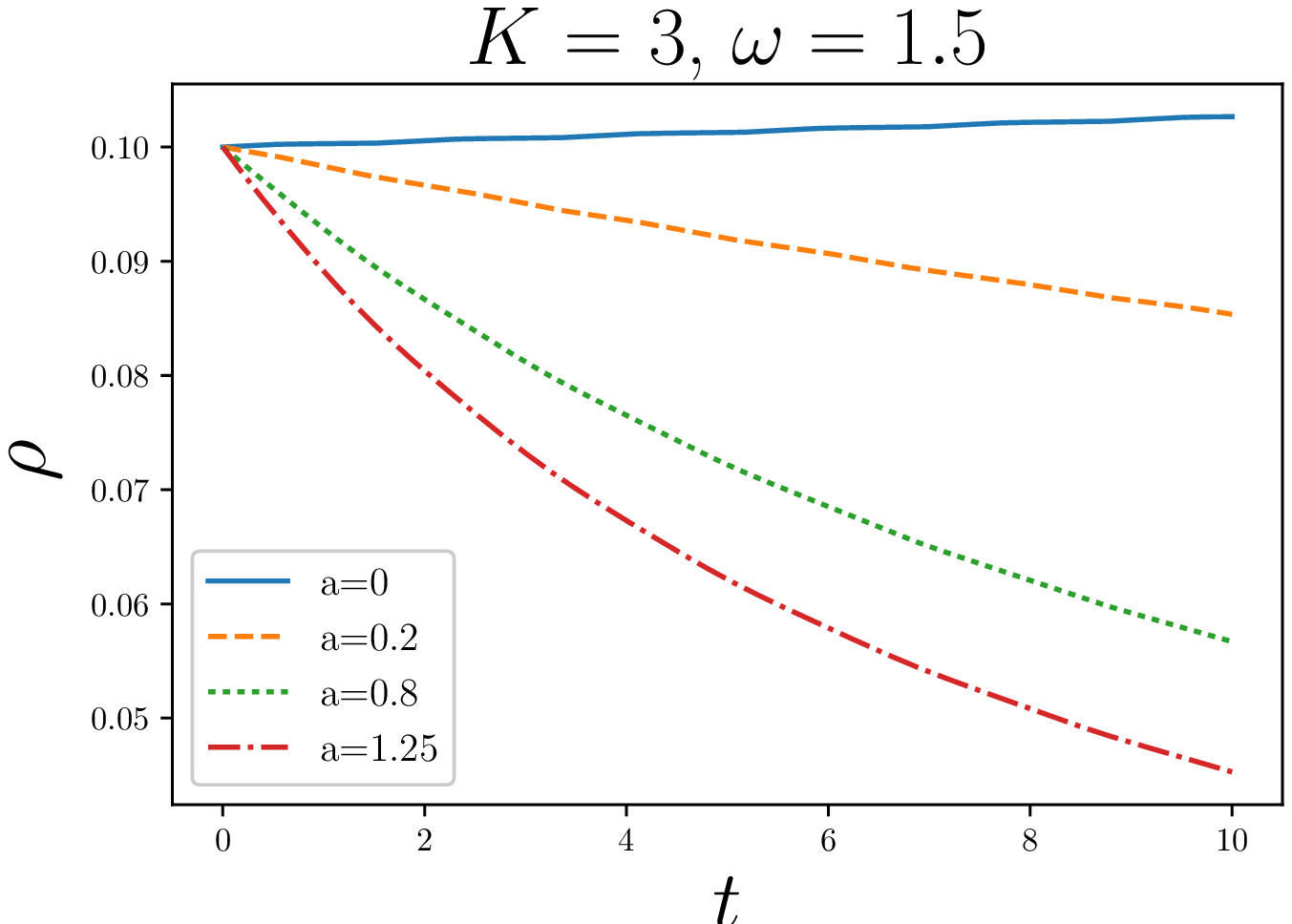}
\end{minipage}\\
\vspace{0.5truecm}
\begin{minipage}[c]{0.5\textwidth}
\includegraphics[width=3.in, height= 2.5in]{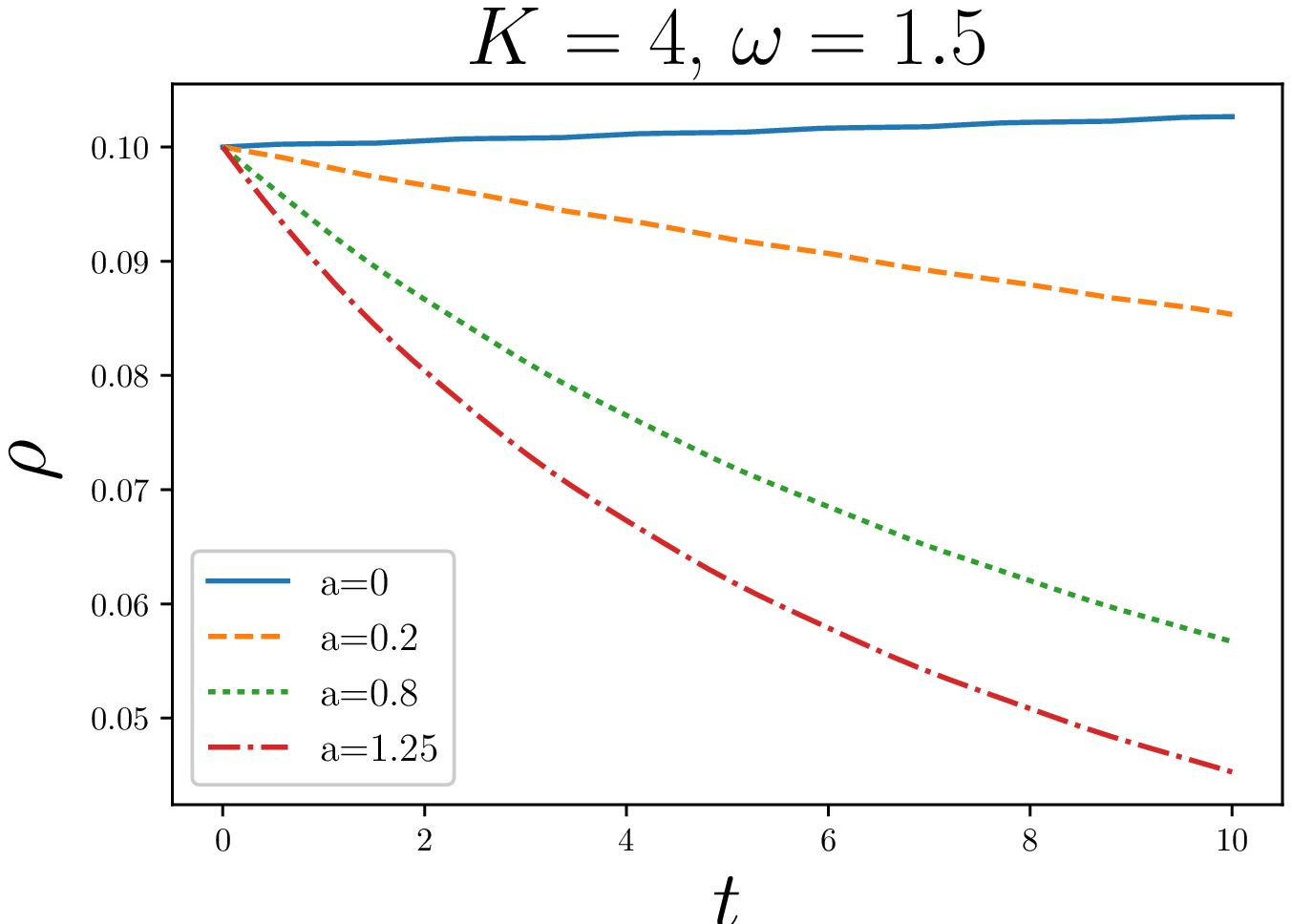}
\end{minipage}%
\begin{minipage}[c]{0.5\textwidth}
\includegraphics[width=3.in, height= 2.5in]{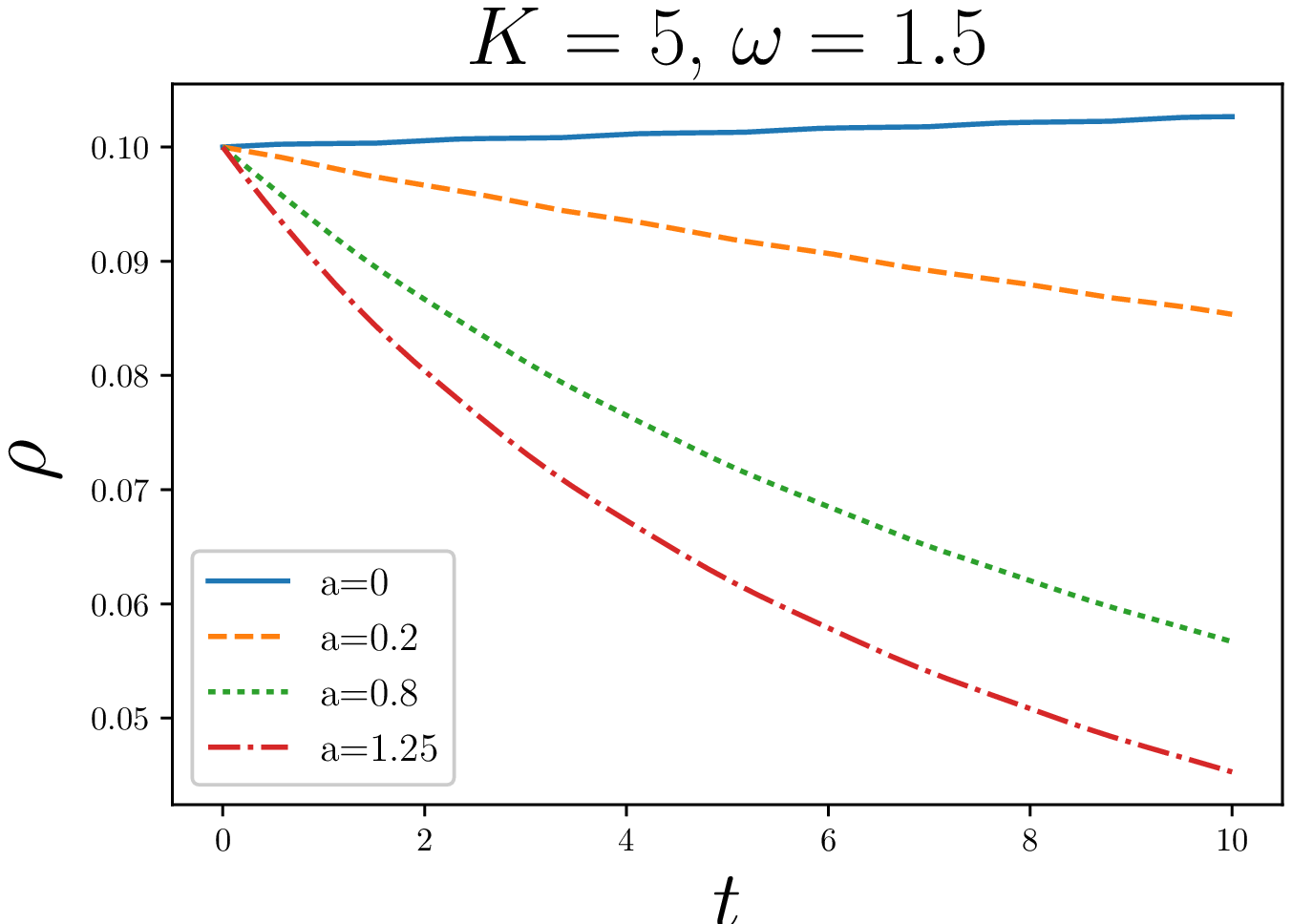}
\end{minipage} 
   \caption{(Color online) Time variation of the electron plasma density $\rho$, for $K=2, 3, 4, 5$ and values of $a$ indicated in the graphs. $\sigma=0.8$, $\omega_0\tau_0=0.8$, $\delta=-0.5$, $\mu=0.5$, $\nu=0.5$, $\gamma=0.18$, $\alpha=0.6$ .}{\label{fig10}}
  \end{figure}

 The electron plasma density $\rho$, plotted in fig. \ref{fig8}, is increasing with time for $a=0$ irrespective of the value of $K$. However, when $a$ is increased in a relatively large range of values, $\rho(t)$ decreases exponentially in time tending to its equilibrium value with an increasingly sharp slope. Evidently, the nonzero values of $a$ chosen for the numerical results just discussed, are large enough and therefore do not enable one appreciate how the electron plasma density $\rho$ changes from its exponentially increasing feature for $a=0$, to an exponentially decreasing feature when $a\neq 0$. In order to earn more insight onto the effects of $a$ on the time variation of $\rho$, in fig. \ref{fig9} and fig. \ref{fig10} we plot $\rho$ versus time keeping the same values of parameters used in fig. \ref{fig8}, but choosing smaller values of $a$. 
    
To be explicit, in fig. \ref{fig9} values of $a$ were chosen small but large enough such that $\rho(t)$ is already decreasing with time. On the contrary the values chosen for $a$ in fig. \ref{fig10}, are closer to zero. The physically relevant insight is that in the range of values of $a$ for which $\rho$ decreases, an increase of $a$ sharpens the slope of the decrease of $\rho$. When $a$ is very small such that $\rho$ is an exponentially increasing function of time, an increase of $a$ will soften the variation of $\rho$. It is quite apparent that in the very small range of values of $a$, $\rho$ is periodically oscillating as it increases exponentially with time (fig. \ref{fig10}). The quantity $a_{th}$ in the graphs of fig. \ref{fig9}, is the characteristic value of the radiative recombination coefficient for which the exponential variation of $\rho$ with time is suppressed. For this value of $a$, the electron plasma density $\rho(t)$ constantly oscillates between two positive extrema and hence is always nonzero, in average.  

\section{Conclusion}
 We have investigated the dynamics of a model for femtosecond laser inscription in a transparent medium with Kerr nonlinearity. This model is an improvement of two recent ones \cite{r19a,r19b}, in which the contribution of electron-hole radiative recombination processes to the plasma generation was neglected. Mathematically the model is represented by a complex Ginzburg-Landau equation with cubic nonlinearity plus a $K$-order nonlinearity accounting for $K$ photon absorption processes, coupled to a time first-order ordinary differential equation with a term quadratic in the electron plasma density, accounting for radiative recombination processes. \par In view of the fact that the laser dynamics in the host material determines the specific regime of operation of the laser during laser micromachining on the material, we considered two distinct regimes of laser dynamics namely CWs and pulses. First addressing CWs and their stability, a linear-stability analysis was carried out following the modulational-instability theory. From this analysis we constructed a global stability map, which enabled us explore parameter regimes in which CW operations could be stable. \par In the nonlinear regime, an ansatz was introduced which aimed at representing the laser as a pulse field with a real amplitude and real phase. With the help of this ansatz the system dynamics was transformed into a set of four first-order ordinary differential equations. Numerical solutions to these equations established the pulse-train structure of the laser amplitude, the power of which was increased by an increase of the radiative recombination coefficient $a$. The electron plasma density was found to increase or decrease exponentially in time with an oscillating amplitude, depending on the order of magnitude of $a$.  
\par Though the model treated in this work is rich, it remains limited for it concerns only few out of a wide range of possible physical contexts. Indeed not all transparent materials oppose a nonlinear optical response of the Kerr type to field propagation, in fact the Kerr response is usually considered as a very weak nonlinearity. There exists transparent materials with stronger optical nonlinear responses, in general their refractive index is a series of powers of the optical field intensity extending beyond the quadratic term. Usually this is also represented by saturable-nonlinearity functions \cite{r23}. Few examples of such materials include doped silica fiber glasses, where the doping with rare-earth ions provides means of controlling the strength of nonlinearity of the transparent host. A study similar to the present one, but taking into consideration the saturable nonlinearity of the host material, will surely enrich current knowledge on the fundamental physics of CW and pulse operations in femtosecond laser micromachining involving transparent materials with non-Kerr optical nonlinearity \cite{r23}. 

\ack
A. M. Dikand\'e thanks the Ministry of Higher Education of Cameroon (MINESUP) for financial assistance, within the framework of the presidential grant designated "Research Modernization Allowances".


\section*{References}
\begin{thebibliography}{99}
\bibitem{r1} F. Dausinger, F. Lichtner and H. Lubatschowski, {\it Femtosecond Technology for Technical and Medical Applications} (Springer, Berlin, 2004).
\bibitem{r2} G. Chryssolouris, {\it Laser machining: theory and practice} ($1^{rst}$ edt., Springer, New York, 1991).
\bibitem{r3} N. B. Dahotre and A. Samant, {\it Laser Machining of Advanced Materials} (CRC Press, Taylor and Francis, 2014).
\bibitem{r4} J. Kr\"uger and W. Kautek, Laser Phys. {\bf 9}, 30 (1999).
\bibitem{r5} M. Stolze, T. Herrmann and J. A. L’huillier, {\it Laser Applications in Microelectronic and Optoelectronic Manufacturing (LAMOM) XXI}, Proceedings of SPIE 9735, 97350O, March 14th (2016).
\bibitem{r6} R. R. Gattass and E. Mazur, Nature Photon. {\bf 2}, April (2008).
\bibitem{r7}A. V. Dostovalov, A. A. Wolf, V. K. Mezentsev, A. G. Okhrimchuk and S. A. Babin, Opt. Expr. {\bf 23}, 32541 (2015).
\bibitem{r8} A. G. Okhrimchuk, V. K. Mezentsev, H. Schmitz, M. Dubov and I. Bennion, Laser Phys. {\bf 19}, 1415 (2009).
\bibitem{r9} S. M. Eaton, G. Cerullo and R. Osellame, {\it Fundamentals of Femtosecond Laser Modification of Bulk Dielectrics}, in: "Femtosecond Laser Micromachining", Topics in Applied Physics (R. Osellame, G. Cerullo and R. Ramponi, eds.) vol 123. (Springer, Berlin, 2012).
\bibitem{r10} A. Couairon and A. Mysyrowicz, Phys. Rep. {\bf 441}, 47 (2007).
\bibitem{r11}H. Shin and D. Kim, Opt. Laser Technol. {\bf 102}, 1 (2018).
\bibitem{r12}P. K. Shukla, J. Lawrence and Y. Zhang, Opt. Laser technol. {\bf 75}, 40 (2015).
\bibitem{r13} Zh. Wang,Y. B. Li, F. Bai, C. W. Wang and Q. Z. Zhao, Opt. Laser Technol. {\bf 81}, 60 (2016).
\bibitem{r14}J. W. Jung and C. M. Lee, {\it Cutting temperature and laser beam temperature effects on cutting tool deformation in laser-assisted machining}, in: Proceedings of IMECS II, pp. 1817–1822 (2009).
\bibitem{r15}H. B. Xu, J. Hu, H. Sheng and Z. C. Du, J. Phys. Conf. Ser. {\bf 633}, 0120901 (2015).
\bibitem{r16}S. M. Eaton, H. Zhang and P. R. Herman, Opt. Expr. {\bf 13}, 4708 (2005).
\bibitem{r17}D. Savastru, R. Savastru, S. Miclos and I. I. Lancranjan, Opt. Laser Eng. {\bf 110}, 288 (2018).
\bibitem{r18}K. Minoshima, A. W. Kowalevicz, I. Hartl, E. P. Ippen and J. G. Fujimoto, Opt. Lett. {\bf 26}, 1516 (2001).
\bibitem{r19}S. Nolte, S., M. Will, J. Burghoff and A. T\"unnermann, J. Mod. Opt. {\bf 51}, 2533 (2004).
\bibitem{r19a}L. Sudrie, A. Couairon, M. Franco, B. Lamouroux, B. Prade, S. Tzortzakis and A. Mysyrowicz, Phys. Rev. Lett. {\bf 89}, 186601 (2002). 
\bibitem{r19b}J. S. Petrovic, V. Mezentsev, H. Schmitz and I. Bennion, Opt. Quant. Electron. {\bf 39}, 939 (2007).
\bibitem{r20}M. Wollenhaupt, A. Assion and T. Baumert, {\it Short and Ultrashort Laser Pulses}, in: "Springer Handbook of Lasers and Optics" (Springer, Berlin, 2012).
\bibitem{r21}T. B. Benjamin and J. E. Feir, J. Fluid Mech. {\bf 27}, 417 (1967).
\bibitem{r22}C. J. Chen, P. K. A. Wai and C. R. Menyuk, Opt. Lett. {\bf 20}, 350 (1995).
\bibitem{r23}A. M. Dikand\'e, J. Voma Titafan and B. Z. Essimbi, J. Opt. {\bf 19}, 105504 (2017).
\bibitem{r24}E. M. Ntongwe and A. M. Dikand\'e, Opt. Quant. Electron. {\bf 51}, 361 (2019). 
\bibitem{r25}A. M. Dikand\'e, Phys. Rev. A{\bf 81}, 013821 (2010).
\bibitem{r26}D. Jr. Jubgang Fandio and A. M. Dikand\'e, J. Opt. Soc. Am. B{\bf 34}, 2721 (2017).
\bibitem{r27}R. D. Dikand\'e Bitha and A. M. Dikand\'e, Phys. Rev. A{\bf 97}, 033813 (2018).
\bibitem{r28}R. D. Dikand\'e Bitha and A. M. Dikand\'e, Eur. Phys. Jour. D{\bf 73}, 152 (2019).
\bibitem{r29}P. Kameni Nteutse, A. M. Dikand\'e and S. Zekeng, Opt. Quantum Elec. {\bf 52}, 313 (2019).
\bibitem{r30}H. Luther, Math. Comp. {\bf 22} 434 (1968).
\bibitem{soto}J. M. Soto-Crespo, N. Akhmediev and G. Town, J. Opt. Soc. Am. {\bf 19}, 234 (2002).
\bibitem{mbieda}F. G. Mbieda Ngomegni, A. M .Dikand\'e and B. Z. Essimbi, Phys. Scrip. {\bf 95}, 025502 (2020). 
\bibitem{gen}J. Lawrence (ed.), {\it Advances in Laser Materials Processing: Technology, Research and Application} (second ed., Elsevier, Amsterdam, 2018).

\end{thebibliography}
\end{document}